\documentclass{ws-mpla}

\begin{document}

\markboth{E. Ruiz Arriola}
{Three pion nucleon coupling constants}

%%%%%%%%%%%%%%%%%%%%% Publisher's Area please ignore %%%%%%%%%%%%%%
\catchline{}{}{}{}{}
%%%%%%%%%%%%%%%%%%%%%%%%%%%%%%%%%%%%%%%%%%%%%%%%%%%%%%%%%%%%%%%%%%%

\title{Three pion nucleon coupling constants~\footnote{
Presented by ERA at 
``Determination of the Fundamental Parameters in QCD'' 
7-12 March 2016,  
Mainz Institute for Theoretical Physics, Johannes Gutenberg University (Mainz,Germanby). \newline 
This work was supported by Spanish Ministerio
de Economia y Competitividad and European FEDER funds (grant
FIS2014-59386-P) and by the Agencia de Innovacion y Desarrollo de
Andalucia (grant No. FQM225).  This work was partly performed under
the auspices of the U.S. Department of Energy by Lawrence Livermore
National Laboratory under Contract No. DE-AC52-07NA27344. Funding was
also provided by the U.S. Department of Energy, Office of Science,
Office of Nuclear Physics under Award No. DE-SC0008511 (NUCLEI SciDAC
Collaboration)}
}

\author{E. RUIZ ARRIOLA, J. E. AMARO}
\address{Departamento de F\'{\i}sica At\'omica, Molecular y Nuclear 
 \\ and Instituto Carlos I de F{\'\i}sica Te\'orica y Computacional. \\
 Universidad de Granada, \\ E-18071 Granada, Spain.
\\
earriola@ugr.es,amaro@ugr.es}

\author{R. NAVARRO P\'EREZ}

\address{Nuclear and Chemical Science Division, Lawrence Livermore National 
Laboratory, Livermore, CA 94551, USA
\\
navarroperez1@llnl.gov}

\maketitle

\pub{Received (Day Month Year)}{Revised (Day Month Year)}

\begin{abstract}
There exist four pion nucleon coupling constants, $f_{\pi^0 pp}$,
$-f_{\pi^0 nn}$, $f_{\pi^+ pn} /\sqrt{2}$ and $ f_{\pi^- np}
/\sqrt{2}$ which coincide when up and down quark masses are identical
and the electron charge is zero. While there is no reason why the
pion-nucleon-nucleon coupling constants should be identical in the
real world, one expects that the small differences might be pinned
down from a sufficiently large number of independent and mutually
consistent data.  Our discussion provides a rationale for our recent determination 
$$f_p^2 = 0.0759(4) \, , \quad f_{0}^2 = 0.079(1) \,, \quad f_{c}^2 =
0.0763(6) \, , $$ based on a partial wave analysis of the $3\sigma$
self-consistent nucleon-nucleon Granada-2013 database comprising 6713
published data in the period 1950-2013.  

\end{abstract}

\ccode{PACS Nos.:03.65.Nk,11.10.Gh,13.75.Cs,21.30.Fe,21.45.+v}

\section{Introduction}	

Four score and a year ago Yukawa brought forth a new theory of
nuclear forces and dedicated to the proposition that protons and
neutrons exchange pions~\cite{Yukawa:1935xg}. But created and
anihilated pions are all not equal. The quest for isospin violations in
particle and nuclear physics has been a permanent goal ever since
Kemmer~\cite{Kemmer:1939zz} invented the concept and generalized the
Pauli principle.  A readable account of the early developpments can be
found in Ref.~\cite{pauli1948meson}.  Actually, the neutral pion was
sought and found because isospin symmetry required it. While the mass
of the pion may be deduced directly from their decays $\pi^0 \to
\gamma \gamma$ and $\pi^\pm \to \mu^\pm + \nu_\mu$ the determination
of the coupling constant to nucleons is more intrincate and needs
further theoretical ellaboration. Although this is not a fundamental
constant of QCD, the pion-nucleon-nucleon coupling constant is the
strong hadronic charge of neutrons and protons, which appear as the
effective constituents of atomic nuclei.

In 1940 Bethe obtained the value $f^2=0.077-0.080$ from the study 
of deuteron properties~\cite{Bethe:1940zz}. Subsequent
determinations based on a variety of processes can be traced from
recent compilations~\cite{deSwart:1997ep,Sainio:1999ba}.  Attempts to
make a microscopic distinction via radiative vertex corrections aiming
at predictive power have been made in the past (see
\cite{henley1969isospin} for pre-QCD account).  At the hadronic level
and within the meson exchange picture there have been many attempts to
determine the many possible causes of isospin breaking, $\rho^0-\omega$ and
$\pi-\eta$ mixing, pion mass differences in the two-pion exchange
interaction, $\pi \gamma$ exchange, etc. (see
e.g. \cite{Miller:1990iz,Miller:2006tv} for post QCD comprehensive
accounts). All these many complications contribute to the belief that
isospin violations in strong interactions remains one of the least
understood issues in the nuclear force.

In this contribution we provide our point of view on this subject and
the elements underlying ongoing work on the determination of the three
coupling constants. Although many of the issues we address here are
well known for the experts after 60 years of NN Partial Wave Analysis fits,
we try in this short account to be pedagogical for the non-experts.
For comprehensive and concise reviews we suggest
Ref.~\cite{deSwart:1997ep,Sainio:1999ba} where the latest NN based 
determination $f^2=0.0750(9)$ is recommended.

\section{Charge Dependent One Pion Exchange (CD-OPE)}

\begin{figure}[tbc]
\begin{center}
\epsfig{figure=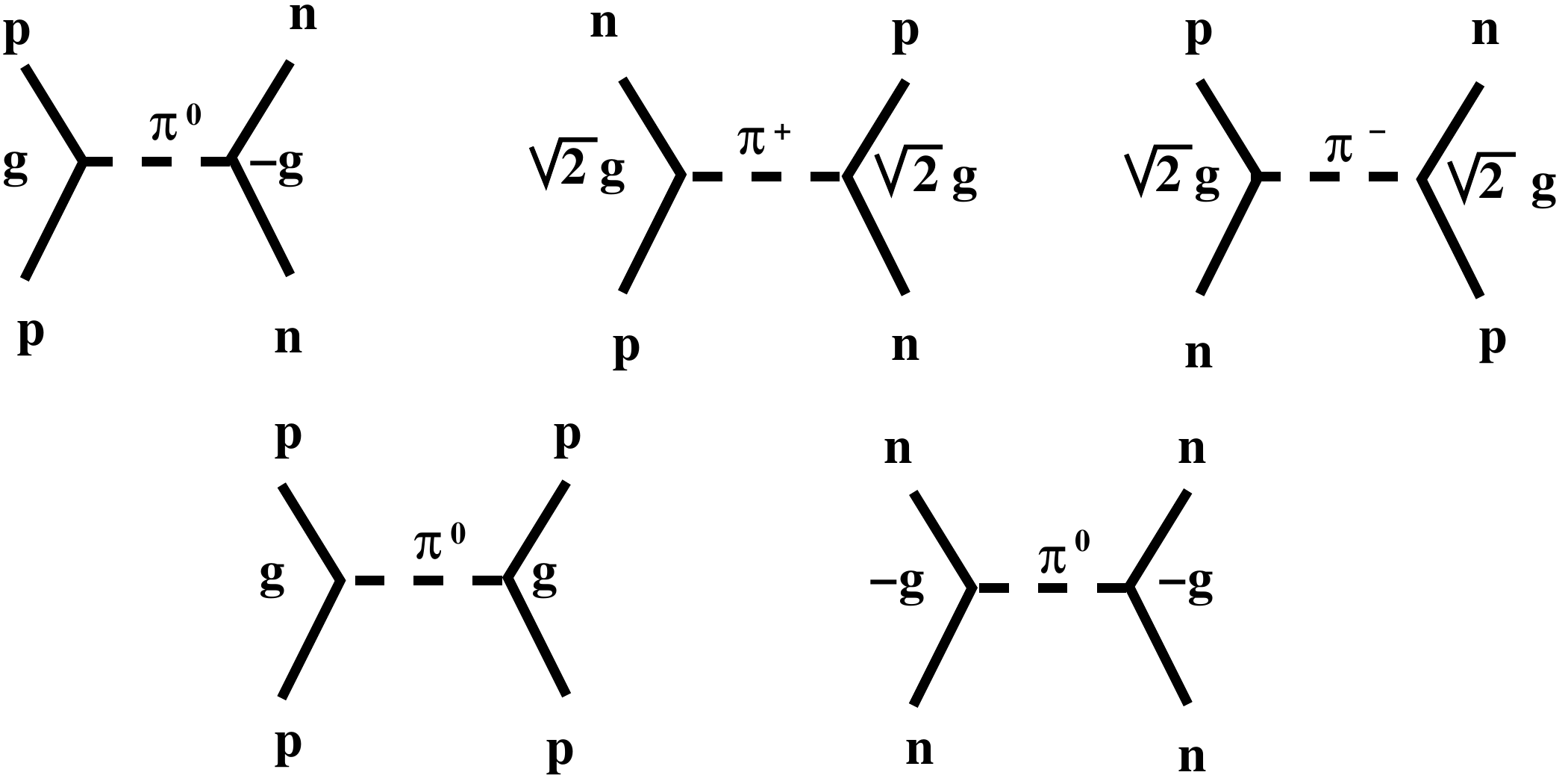,width=0.8\linewidth}  
\end{center}
\caption{Feynman diagrams contributing to the charge dependent 
one pion exchange interaction. The couplings are assumed to be 
in the isospin limit $g=g_{\pi^0 pp}= g_{\pi^\pm n p}/\sqrt{2} =
-g_{\pi^0 n n}$.}
\label{Fig:piNN-OPE}
\end{figure}

The coupling constant is rigurously defined as the $\pi NN$ three
point vertex function when all three particles are on the mass shell,
a condition that cannot generally be satisfied for real momenta.  At
the hadronic level the pion-nucleon-nucleon vertex is described by the
Lagrangian (we use the standard convention~\cite{Dumbrajs:1983jd})
\begin{eqnarray}
  {\cal L} = \sqrt{4\pi} \frac{f_{e,ab}}{m_{\pi^+}} \bar B_a \gamma^\mu \gamma_5 B_b \partial_{\mu}
 \phi_e \, , 
\end{eqnarray}
where $\bar B_a$, $B_b$ and $\phi_e$ are antibaryon, baryon and
pseudoscalar meson field respectively and $\gamma^\mu$ and $\gamma_5$
are Dirac matrices.  This yields four possible vertices
\begin{eqnarray}
p \to \pi^+ n \, ,  \quad n \to \pi^- p \, ,  \quad p \to \pi^0 p \, , \quad n \to \pi^0 n \, ,
\end{eqnarray}
so their amplitudes are,  
\begin{eqnarray}
{\cal A} (p \to \pi^+ n) = f_{\pi^+ np } \bar u_n (p_n, s_n)
\gamma^\mu \gamma_5 u_p (p_p, s_p) q_\mu \, , 
\end{eqnarray}
and so on. The relevant relationships between the pseudoscalar pion
coupling constants, $g_{\pi NN}$, and the pseudovector ones, $f_{\pi
  NN}$, are given by
\begin{eqnarray}
\frac{g_{\pi^0 pp}}{\sqrt{4\pi}} = 
       \frac{2M_p}{m_{\pi^\pm}}  f_{\pi^0 pp} \,, \quad 
%      = 180.779 f^2_{\pi^0 pp} \\ 
\frac{g_{\pi^0 nn}}{\sqrt{4\pi}} = 
      \frac{2M_n}{m_{\pi^\pm}} f_{\pi^0 nn} \, , \quad 
%      =  181.278 f^2_{\pi^0 nn}\\
\frac{g_{\pi^\pm pn}}{\sqrt{4\pi}} = \frac{M_p+M_n}{m_{\pi^\pm}}  
 f_{\pi^\pm p n} \, .  
%= 181.029 f^2_{\pi^\pm pn} \; .
\end{eqnarray}
with $M_p=938.272$ MeV the proton mass, $M_n=939.566$ MeV the neutron
mass, and $m_{\pi^\pm}=139.570$ MeV the mass of the charged pion.
From these vertices one may obtain the NN scattering amplitude to
lowest order in perturbation theory, see Fig.~\ref{Fig:piNN-OPE}. For
instance, the unpolarized differential nn-cross section
reads
\begin{eqnarray}
\frac{d \sigma_{nn}}{dt} &=&  \frac{g_{\pi^0 nn}^4 }{32 \pi s (s- 4 M_n^2)}
\left[ \frac{t^2}{(t-m_{\pi^0}^2)^2}-\frac{2 (\frac{t}2-M_n^2)^2 - (\frac{s}2-M_n^2)^2- 4 M_n^2}{(t-m_{\pi^0}^2)(u-m_{\pi^0}^2)} \right] \nonumber \\ 
&+& (u \leftrightarrow t) \, ,  
\end{eqnarray}
where $s = 4 (p^2 + M_n^2)$, 
$t=-q^2= - 4 p^2 \sin(\theta/2)^2 < 0 $ and $s+t+u= 4M_n^2$
with $(p,\theta)$ 
the CM momentum and scattering angle respectively. This is perhaps the most
straightforward way of checking OPE from data, i.e. extrapolating
$(t-m_{\pi^0}^2)^2 d\sigma/dt $ from the physical $t<0$ kinematics to
the unphysical limit $t \to m_{\pi^0}^2$. Unfortunately, the Born
approximation violates elastic unitarity and there are many ways to
restore it.  For reasons to be explained below we use the
phenomenological potential approach and a partial wave expansion
analysis~\cite{signell1969nuclear}. The concept of a potential is essentially
non-relativistic and the procedure to obtain
it is to match perturbatively the non-relativistic quantum mechanical
potential to the same scattering amplitude obtained in quantum field theory
for the direct term, namely
\begin{eqnarray}
f_{\rm QFT}^{\rm Born} (\theta, E)= -\frac{2 \mu}{4\pi} \int
d^3 x e^{-i \vec k' \cdot \vec x} V(\vec x, \vec p) e^{i \vec k\cdot
  \vec x} \, , 
\end{eqnarray}
where the on-shell condition is understood $k'=k=p$.  This already
incorporates an ambiguity, since one may add terms which vanish on the
mass-shell. In the static limit of heavy nucleons the ambiguity
dissapears and the potential deduced from field theory becomes local,
which is of great advantage since we obtain a Schr\"odinger equation.
The CD-OPE potential in the pp and np channels so obtained reads
\begin{eqnarray}
 V_{pp \to pp}(r) &=& f^2_{\pi^0 pp} V_{m_{\pi^0}}(r),  \\
 V_{np \to np}(r) = V_{pn \to pn}(r) 
&=& -f_{\pi^0 nn}f_{\pi^0 pp}V_{m_{\pi^0}}(r) \\ 
V_{pn \to np}(r) = V_{np \to pn}(r) &=& f_{\pi^-pn} f_{\pi^+np} \, 
 V_{m_{\pi^\pm}}(r) \\ 
V_{nn \to nn}(r) &=& f^2_{\pi^0 nn} V_{m_{\pi^0}}(r),  
 \label{eq:BreakIsospinOPE}
\end{eqnarray}
where $V_{m,\rm OPE}$ is given by
\begin{equation}
V_{m, \rm OPE}(r) =
\left(\frac{m}{m_{\pi^\pm}}\right)^2\frac{1}{3}\frac{e^{-m r}}{ r}
\left[ {\mathbf \sigma}_1\cdot\mathbf{\sigma}_2 + \left( 1 +
  \frac3{mr} + \frac{3}{(mr)^2}\right) S_{12} \right] \, ,
\end{equation}
with $\sigma_1$ and $\sigma_2$ the single nucleon Pauli matrices and
$S_{12}= 3 \sigma_1 \cdot {\bf \hat r} \sigma_2 \cdot {\bf \hat r}
-\sigma_1 \cdot \sigma_2 $ the tensor operator. Using the standard
notation we make the identifications
\begin{eqnarray}
f_p^2 = f_{\pi^0 pp} f_{\pi^0 pp} \, , \quad 
f_0^2 = -f_{\pi^0 nn} f_{\pi^0 pp} \, , \quad 
2 f_c^2 = f_{\pi^-pn} f_{\pi^+np} \, . 
\end{eqnarray}
In many derivations of the OPE potential some emphasis is placed on
the contact and singular piece which is proportional to a Dirac delta
function located at the origin. We omit them here, since they will
play no role in our subsequent discussion.

\section{The static nuclear potential}

The saturation property of nuclei suggests that there is an
equilibrium distance between two nucleons. While one talks about
nuclear forces, the truth is that they have never been measured
directly in experiment. From a purely classical viewpoint, this would
require to pull two nucleons apart at distances larger than their size
and measure the necessary force, similarly as Coulomb and Cavendish
did for the electric and gravitational forces 250 years ago. For such
an ideal experiment the behaviour of the system at shorter distances
would be largely irrelevant. This situation would naturally occurr if
nucleons were truely infinitely heavy sources of baryon charge. From a
fundamental point of view the nucleon force is defined from the static
energy between two nucleons which in QCD are made of three quarks and
any number of quark-antiquark pairs and gluons.  Nucleons and pions
are composite and extended particles which can be characterized by any
gauge invariant combination of interpolating fields with the proper
quantum numbers. This generates some ambiguity except for heavy
quarks.  In any case the static energy reads
\begin{eqnarray}
E_{NN}(R) = 2 M_N + \sum_{q \in N,q' \in N'} V_{q,q'} ( \vec x_q - \vec x_{q'} ) \, . 
\end{eqnarray}
Here, the quarks in each nucleon are located at the same point $\vec
x_q=\vec R/2$ and $\vec x_{q'}=-\vec R/2$. On the lattice static
baryon sources have been placed at a fixed
distance~\cite{Aoki:2011ep,Aoki:2013tba}, and in fact there exist
lattice calculations addressing the pion-nucleon coupling
constant. Momentum dependence of the strong form factor and coupling
constant determination in lattice QCD gives $g_{\pi NN} \sim 10-12
$~\cite{Alexandrou:2007xj} and more recently $g_{\pi NN} \sim 13(1) $
for $m_\pi \sim 560 {\rm MeV}$ \cite{Erkol:2008yj} using QCD sum
rules.  On the lattice in the quenched approximation it has been found
that for a pion mass of $m_\pi=380$ MeV the value $g_{\pi
  NN}^2/(4\pi)=12.1\pm 2.7$ which is encouraging~\cite{Aoki:2009ji}.
These are courageous efforts which still are far from the $1\%$
accuracy needed for witnessing isospin breaking in the couplings (see below).

Some intuition may be gathered from a chiral quark model picture.  At
scales above the confinement radius we expect exchange of purely
hadronic states and at long distances the OPE mechanism will
dominate~\cite{Manohar:1983md}, but because the quarks in each nucleon
are on top of each other, they will contribute coherently to the
couplings. Of course, $u$ and $d$ quarks are not heavy, but
spontaneous chiral symmetry breaking will provide them with a
constituent mass $M_0$ which will give a total mass $M_q=M_0+m_q$.
Thus, neglecting em contributions, the proton and neutron masses are
$M_p= 3 M_0 + 2 m_u + m_d$ and $M_n= 3 M_0 + m_u + 2 m_d$, so that
$M_p-M_n=m_d-m_u$, a reasonable value. Within this picture, relative
corrections $\delta g/g$ at the nucleon or quark level are the
same. The $\pi q q $ coupling is the residue of the Bethe-Salpeter
equation at the pion pole. In a model where the pion is composite such
as the NJL model (see e.g. \cite{RuizArriola:2002wr} for a review) one
has for $m_u,m_d \ll M_0$ ($\Lambda$ is the NJL cut-off)
\begin{eqnarray}
\frac{\delta g}{g}\Big|_{\pi q \bar q} = A (M_0,\Lambda) \frac{m_q+m_{\bar q}}{M_0} + \dots 
\end{eqnarray}
with $A(M_0,\Lambda)>0$ so the coupling to
the pion grows with the quark mass. Thus  
\begin{eqnarray}
\frac{\delta g}{g}\Big|_{\pi^0 n n} > \frac{\delta g}{g}\Big|_{\pi^+ n p} > \frac{\delta g}{g}\Big|_{\pi^0 pp}  \, . 
\label{eq:qm-order}
\end{eqnarray}

\section{Scattering and analytical properties}

Unfortunately we cannot carry out the classical Cavendish experiment
for nucleons in the laboratory. We may instead analyze the about 8000
pp and np scattering data collected in accelerators below LAB energy
$350 {\rm MeV}$ in the period 1950-2013. Quantum mechanically
wave-particle duality implies a finite wavelength resolution for a
given relative momentum, $\Delta r \sim 1/p$. This will allow to
sample the potential at coarse grained distance scales, $r_n\sim n
\Delta r$.  Fluctuations of the interaction below that scale will be
unobservable and manifest as correlations among the interaction at the
sampled points $V(r_n)$.  The ``force'' is defined as the average
change of $V(r_n)$ over the resolution scale $\Delta r$, for instance 
$ 
F(r_n) =- (V(r_n + \Delta r/2)- V(r_n - \Delta r/2))/\Delta r 
$. 
\begin{figure}[tbc]
\begin{center}
\epsfig{figure=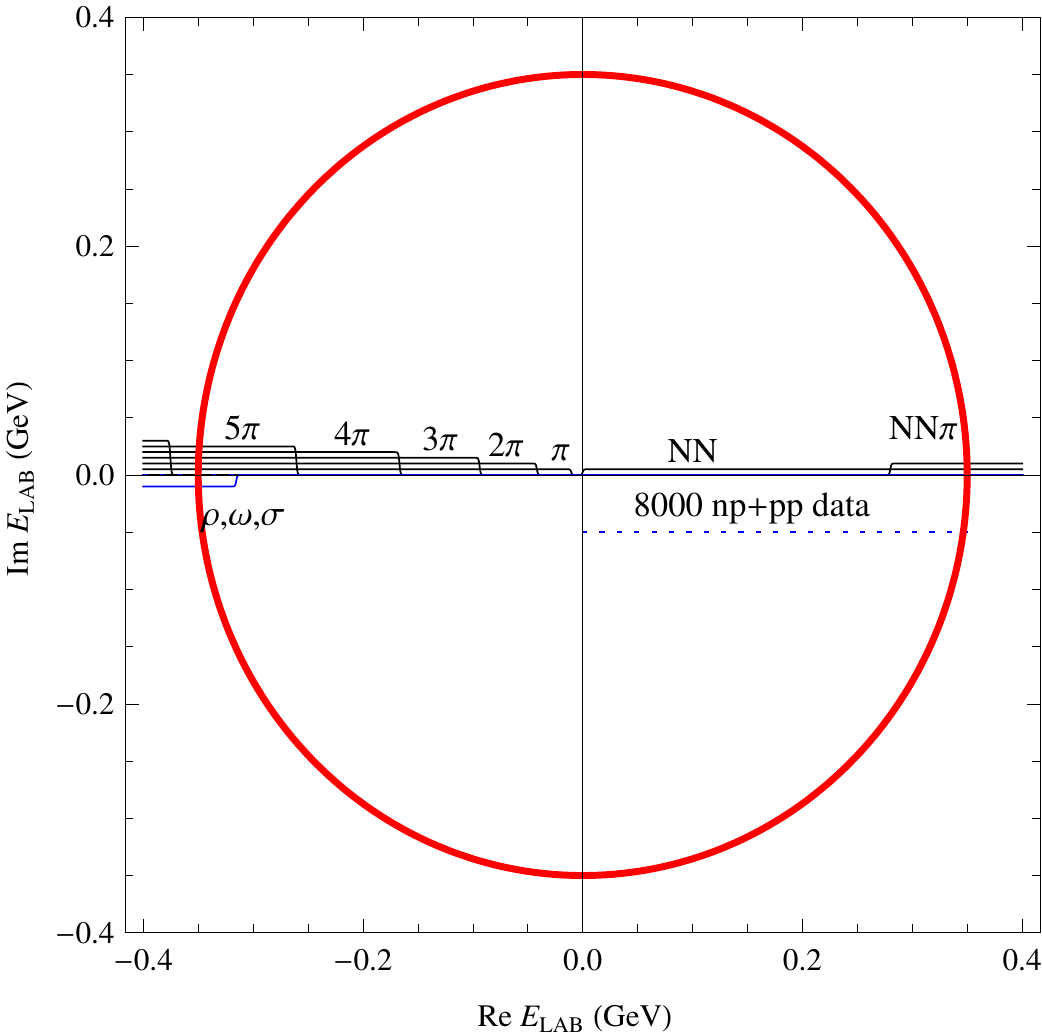,width=0.45\linewidth}
\epsfig{figure=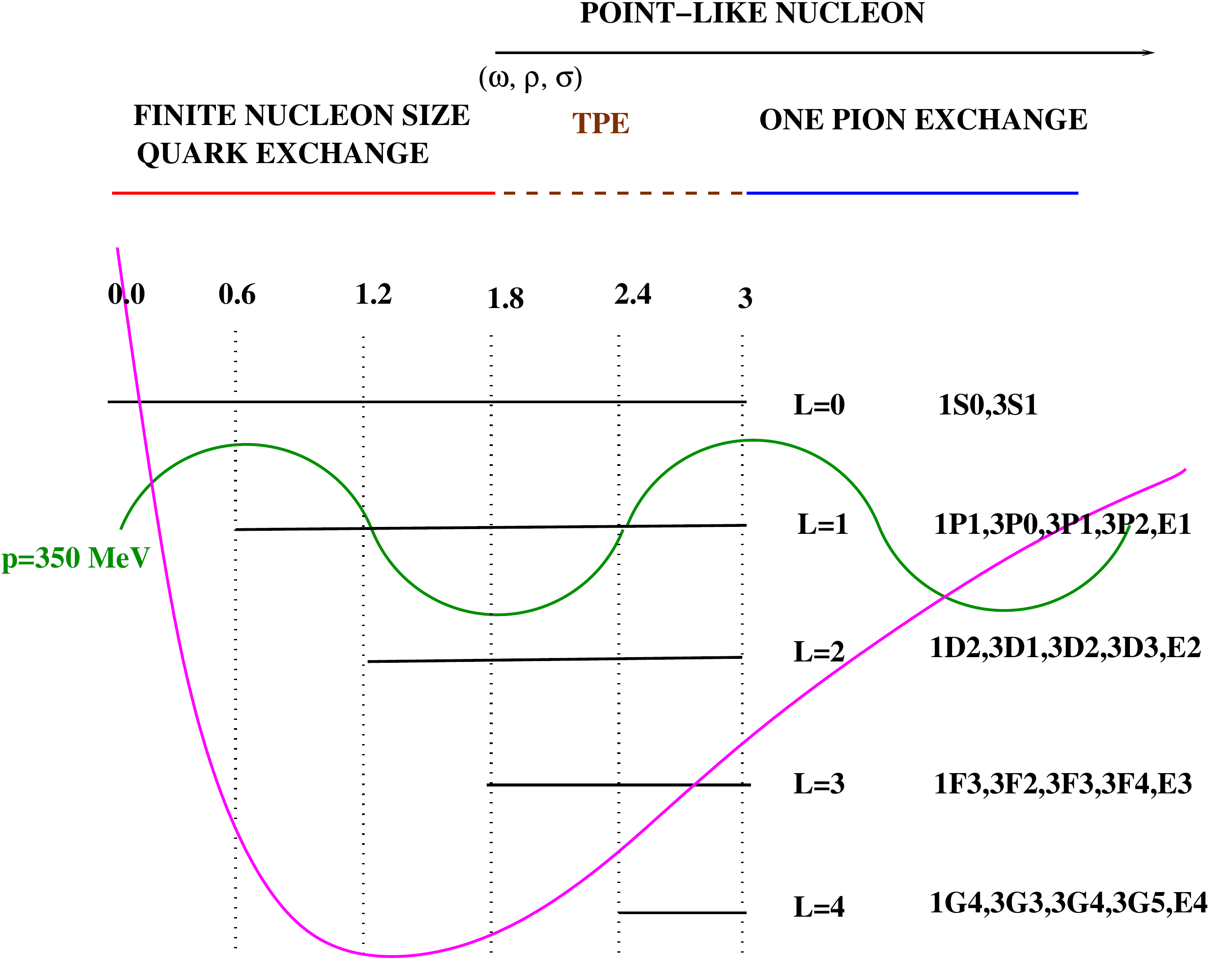,height=0.5\linewidth,width=0.45\linewidth}  
\end{center}
\caption{Two complementary views of the anatomy of the NN interaction.
  Left panel: The LAB energy complex energy plane, showing the partial
  waves left cut structure due to multipion 
%  $T_{\rm LAB}^{\rm left}= ( \dots,  -375.3,-260.6,-166.8,-93.8,-41.7,-10.4) {\rm MeV}$
and $\sigma,\rho,\omega$ exchange and the right cut structure due to
pion production. Right panel: NN Potential as a function of distance,
compared with a free wave, $\sin(pr)$, with $p=2 k_F$, the relative
momentum corresponding to back-to-back scattering in nuclear matter;
the most energetic process inside a heavy nucleus.}
\label{Fig:TLAB-complex}
\end{figure}

The NN scattering amplitude has five independent complex components
which are a function of energy and scattering angle,
\begin{eqnarray}
M 
&=& a
+ m (\mathbf{\sigma}_1\cdot\mathbf{n})(\mathbf{\sigma}_2\cdot\mathbf{n}) 
+ (g-h)(\mathbf{\sigma}_1\cdot\mathbf{m})(\mathbf{\sigma}_2\cdot\mathbf{m}) 
\nonumber \\ 
&+& 
(g+h)(\mathbf{\sigma}_1\cdot\mathbf{l})(\mathbf{\sigma}_2\cdot\mathbf{l})  
+ c (\mathbf{\sigma}_1+\mathbf{\sigma}_2)\cdot\mathbf{n} \, . 
\end{eqnarray}
We use the three unit vectors ($\mathbf{k}_f$ and $\mathbf{k}_i$
are relative final and initial momenta), 
\begin{eqnarray}
\mathbf{l}&=&  \frac{\mathbf{k}_f+\mathbf{k}_i}{|\mathbf{k}_f+\mathbf{k}_i|} \, , 
 \qquad 
 \mathbf{m}=  \frac{\mathbf{k}_f-\mathbf{k}_i}{|\mathbf{k}_f-\mathbf{k}_i|} \, , \qquad 
 \mathbf{n}=  \frac{\mathbf{k}_f \wedge \mathbf{k}_i}{|\mathbf{k}_f \wedge\mathbf{k}_i|} \, . 
\end{eqnarray}
For this amplitude the partial wave expansion in this case reads
\begin{eqnarray}
 M^s_{m_s',m_s}(\theta) &=&  \frac{1}{2ik} \sum_{J,l',l}\sqrt{4\pi(2l+1)}Y^{l'}_{m_s'-m_s}(\theta,0) \nonumber \\
      &\times& C^{l',S,J}_{m_s-m_s',m_s',m_s}i^{l-l'}(S^{J,S}_{l,l'}-\delta_{l',l}) C^{l,S,J}_{0,m_s,m_s},
 \label{eq:MmatrixPartialWaves}
\end{eqnarray}
where $S$ is the unitary coupled channel S-matrix, and the $C's$ are
Clebsch-Gordan coefficients. One has that ${\bf S}^{JS} = ({\bf
  M}^{JS} - i {\bf 1})({\bf M}^{JS} + i {\bf 1})^{-1} $ with $({\bf
  M}^{JS})^\dagger = {\bf M}^{JS}$ a hermitian coupled channel matrix
(also known as the K-matrix). At the level of partial waves the
multipion exchange diagrams generate left hand cuts in the complex
s-plane, which come in addition to the NN elastic right cut and the
$\pi NN$, $ 2\pi NN$ etc., pion production cuts. At low energies for $
|p| \le m_\pi/2$ we have~\cite{PavonValderrama:2005ku}
\begin{eqnarray}
 p^{l+l'+1} M_{l,l'}^{JS} (p) = -(\alpha^{-1})^{JS}_{l,l'} + \frac12  (r)^{JS}_{l,l'}p^2+  (v)^{JS}_{l,l'}p^4 + \dots
\label{eq:ERE-coup}
\end{eqnarray}
which is the coupled channels effective range expansion. We sketch the
situation in Fig.~\ref{Fig:TLAB-complex} (left panel) where a possible
contour for a dispersion relations study is also depicted. When the
cuts are explicitly taken into account there still remains the
question on the number of subtraction constants (see
e.g. Ref.~\cite{Oller:2014uxa} and references therein). While one can
pursue such an analysis, comparison with experimental data goes beyond
just partial waves and generates complications coming from long range
effects, which are most efficiently treated in coordinate space within
the potential approach preferred by nuclear physicists. The magnetic
moment interaction which decreases as $1/r^3$ is crucial to describe
the data but remains a challenge in momentum space. Even in the
friendly coordinate space these terms usually need summing about 1000-2000
partial waves. This is a serious bottleneck for any analysis aiming at
a direct comparison to experimental data. Fortunately, any pion cut
generates a contribution to the potential which at long distances
falls off as $e^{-n m_\pi r}$. The potential in coordinate space,
sketched in Fig.~\ref{Fig:TLAB-complex} (right panel), has the same
analytical structure (left panel) but permits incorporating these
otherwise difficult long distance effects by simply solving the
Schr\"odinger equation and using the sampled, coarse grained potential
$V(r_n)$, as fitting parameters themselves.

\section{An elementary determination of the coupling constant}

One motivation to study the NN force is to apply it to nuclear
structure and nuclear reactions. Many studies are conducted with this
application in mind, when not specifically designed to produce a
potential friendly to some preferred computational many-body method.
This introduces a bias and hence a systematic error in the analysis of
nuclear forces which is often forgotten~
. Thus, fits rarely go much
beyond the pion production region, since the inelastiticy becomes
important above 350 MeV LAB energy and the addition of $\pi NN$
channel does not improve the description.

On the other hand, being composite particles made out of quarks and
gluons, nucleons have a finite size which can be determined by a
variety of methods, mostly by electron and neutrino scattering. To
what extent this finite size is relevant for nuclear structure
calculations is not completely obvious, but in practice nucleons in
nuclear physics are treated as elementary and pointlike. Excitations
such as the $\Delta$ resonance are explicitly included themselves as
elementary point like particles as well. The most dense known nuclear
system is nuclear matter where an average equilibrium separation
distance between nucleons is $d_{NN} \sim 1.8$fm. Thus, nuclear
binding is obviously related to this mid-range distance. On the other
hand the pion production threshold happens at LAB energy $\sim 2 m_\pi
=280 {\rm MeV}$.

\begin{figure}[tbc]
\begin{center}
\epsfig{figure=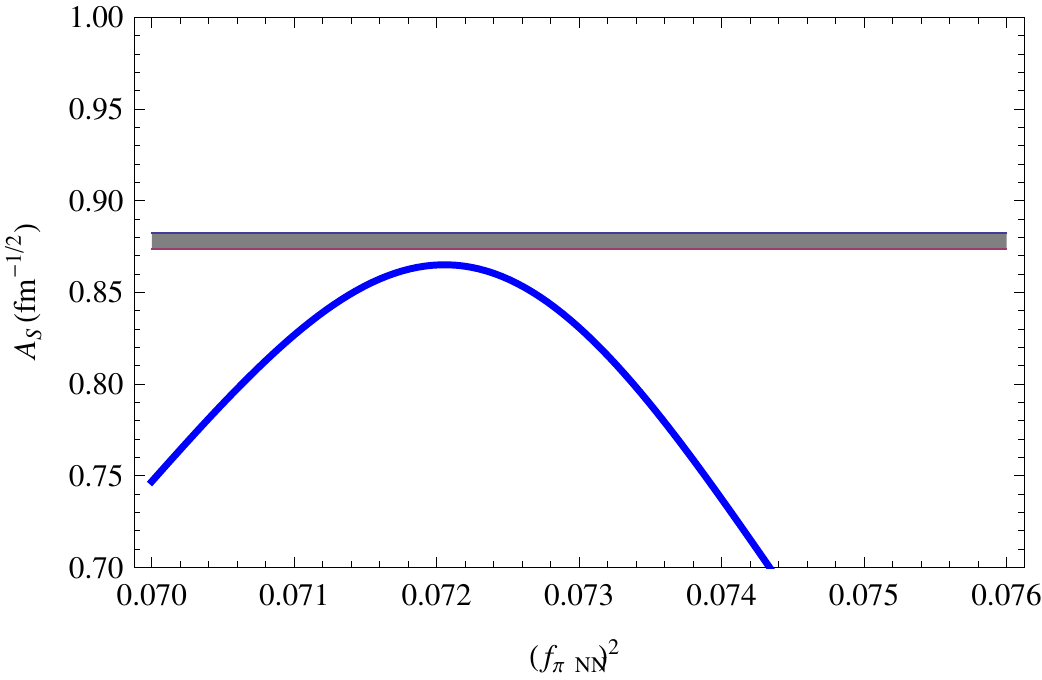,width=0.45\linewidth}  
\epsfig{figure=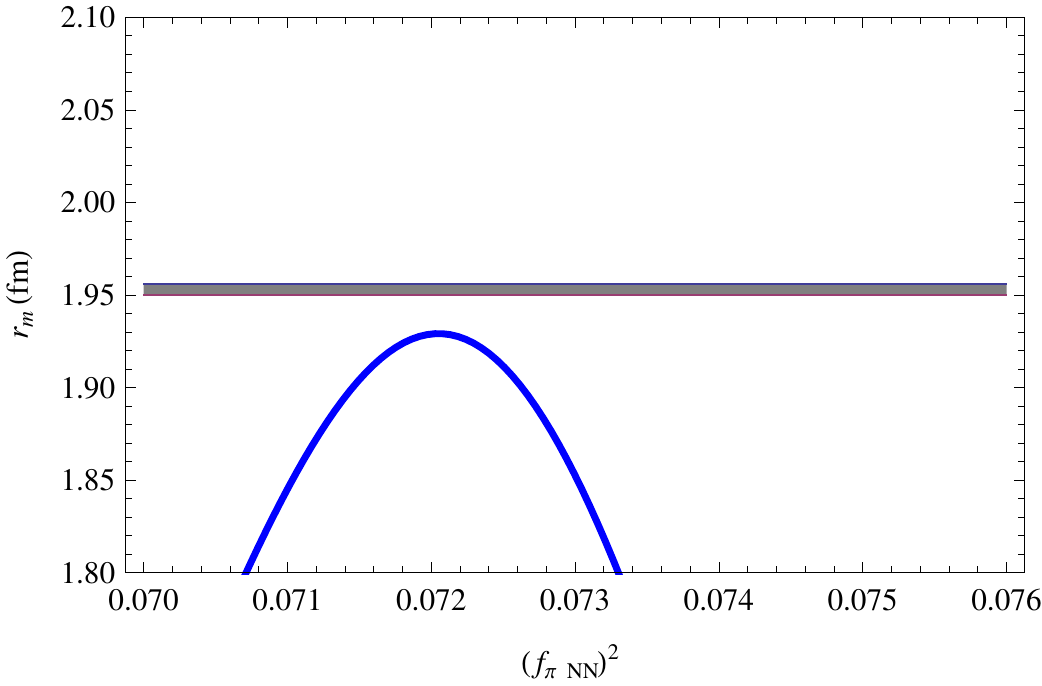,width=0.45\linewidth}  
\epsfig{figure=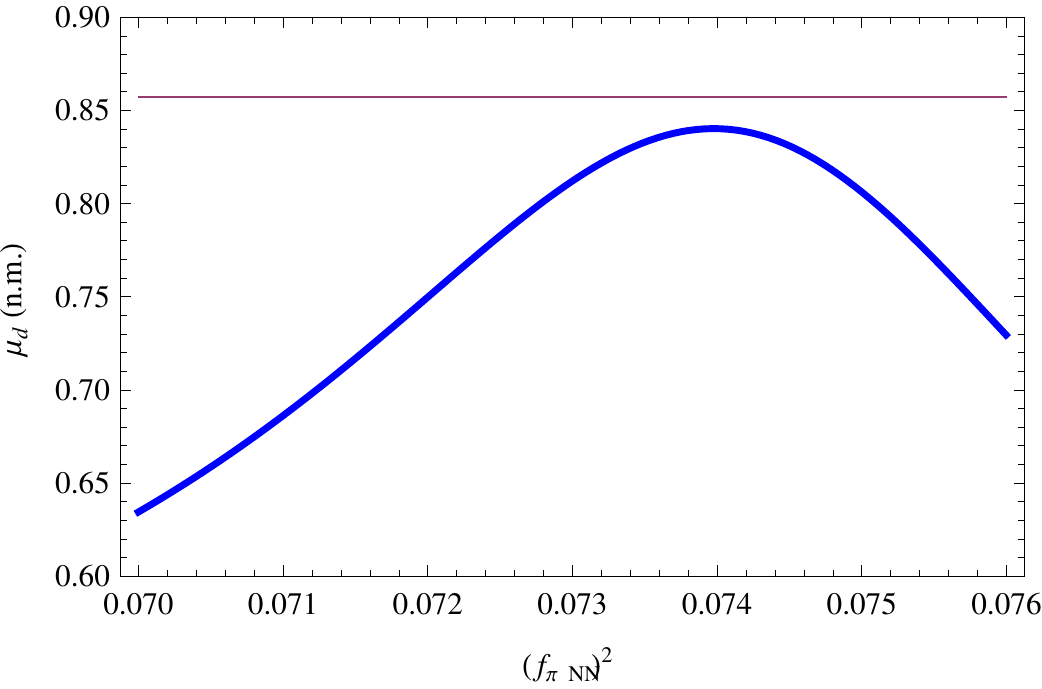,width=0.45\linewidth}  
\epsfig{figure=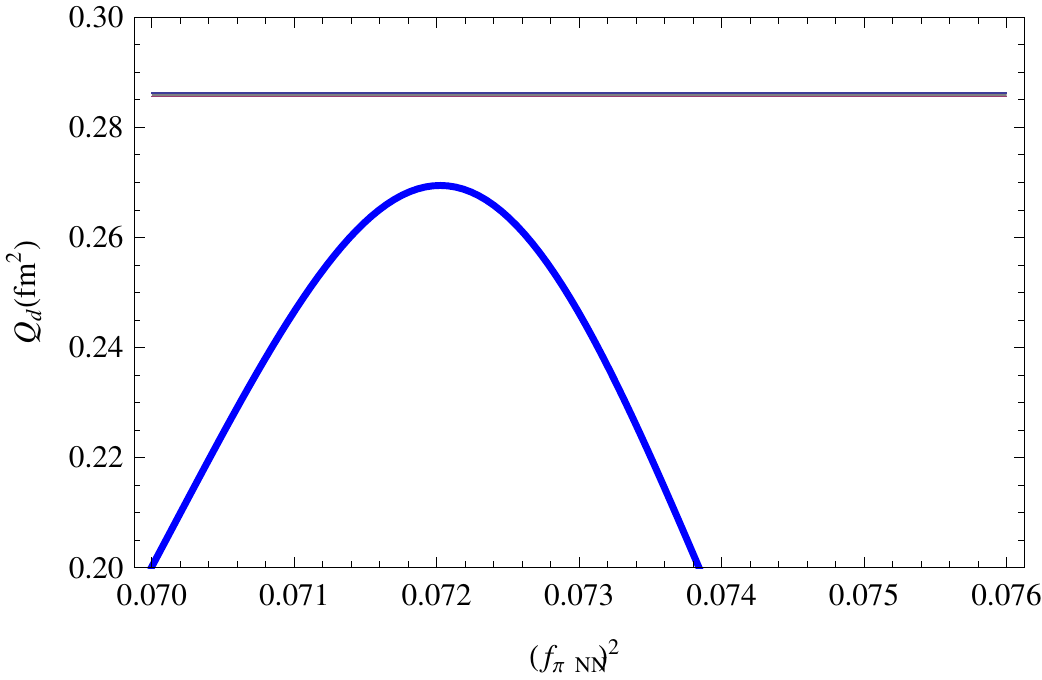,width=0.45\linewidth}  
\end{center}
\caption{Renormalized OPE deuteron properties using the physical
  binding energy, average pion mass and asymptotic D/S ratio
  $\eta=0.0256(4)$ as a function $f_{\pi NN}^2$ compared with experiment.}
\label{Fig:piNN-OPE-deuteron}
\end{figure}

An extreme situation corresponds to assume that the elementarity
radius $r_e$  is arbitrarily small. The lightest nucleus, the deuteron,
is a bound np system with $J^C=1^+$, and corresponds to a
$^3S_1-^3D_1$ mixed state. The tensor part of the OPE potential
diverges as $1/r^3$ as already pointed out by
Bethe~\cite{Bethe:1940zz}. Thus, the deuteron equations are singular
at short distances, but they can be
renormalized~\cite{PavonValderrama:2005gu} by imposing physical
renormalization conditions. In Fig.~\ref{Fig:piNN-OPE-deuteron} we
show results for a number of renormalized observables and {\it
  requiring} the state to be normalizable at short distances, showing
that $f^2 = 0.072-0.074$, not far from Bethe's venerable value.

\begin{figure}[tbc]
\begin{center}
\epsfig{figure=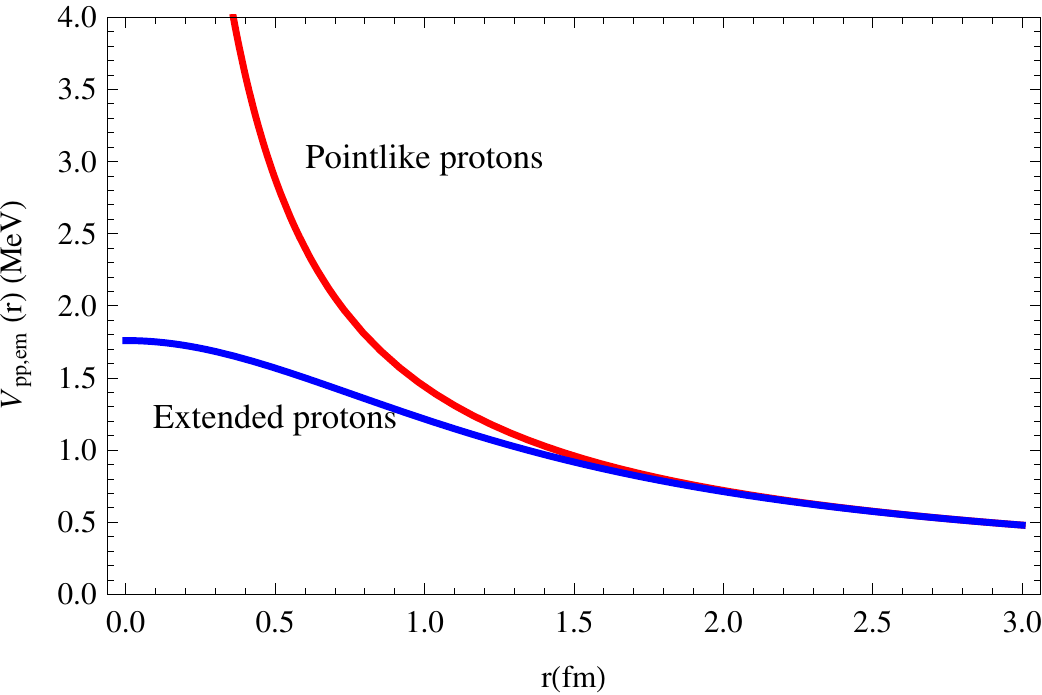,width=0.45\linewidth} 
\hskip.5cm  
\epsfig{figure=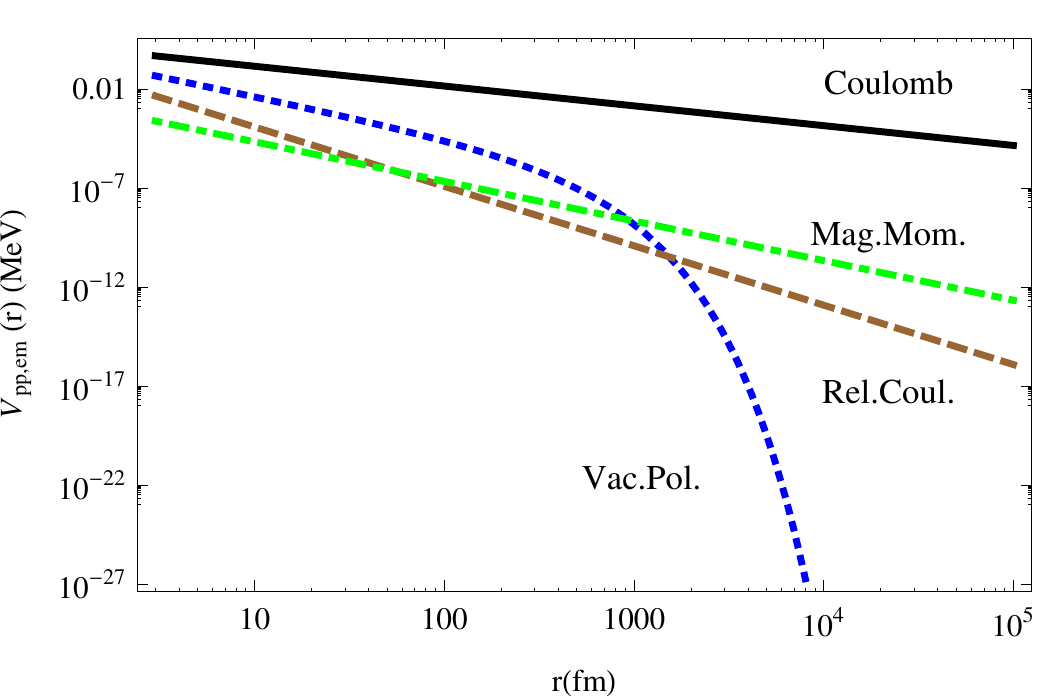,width=0.47\linewidth}  
\end{center}
\caption{Left panel: Proton-proton electrostatic interaction as a
  function of the separation distance. We compare point like protons
  with the charge density deduced from the electric form factor. 
Right panel: : Electromagnetic interactons (note the
  log-log scale): We show the Coulomb interaction, Magnetic Moments,
  Relativistic Coulomb, Vacuum polarization.}
\label{Fig:Vlong}
\end{figure}

\section{NN interaction and Effective elementarity of the Nucleon}

\begin{figure}
\begin{center}
\includegraphics[height=4.1cm,width=4cm,angle=-90]{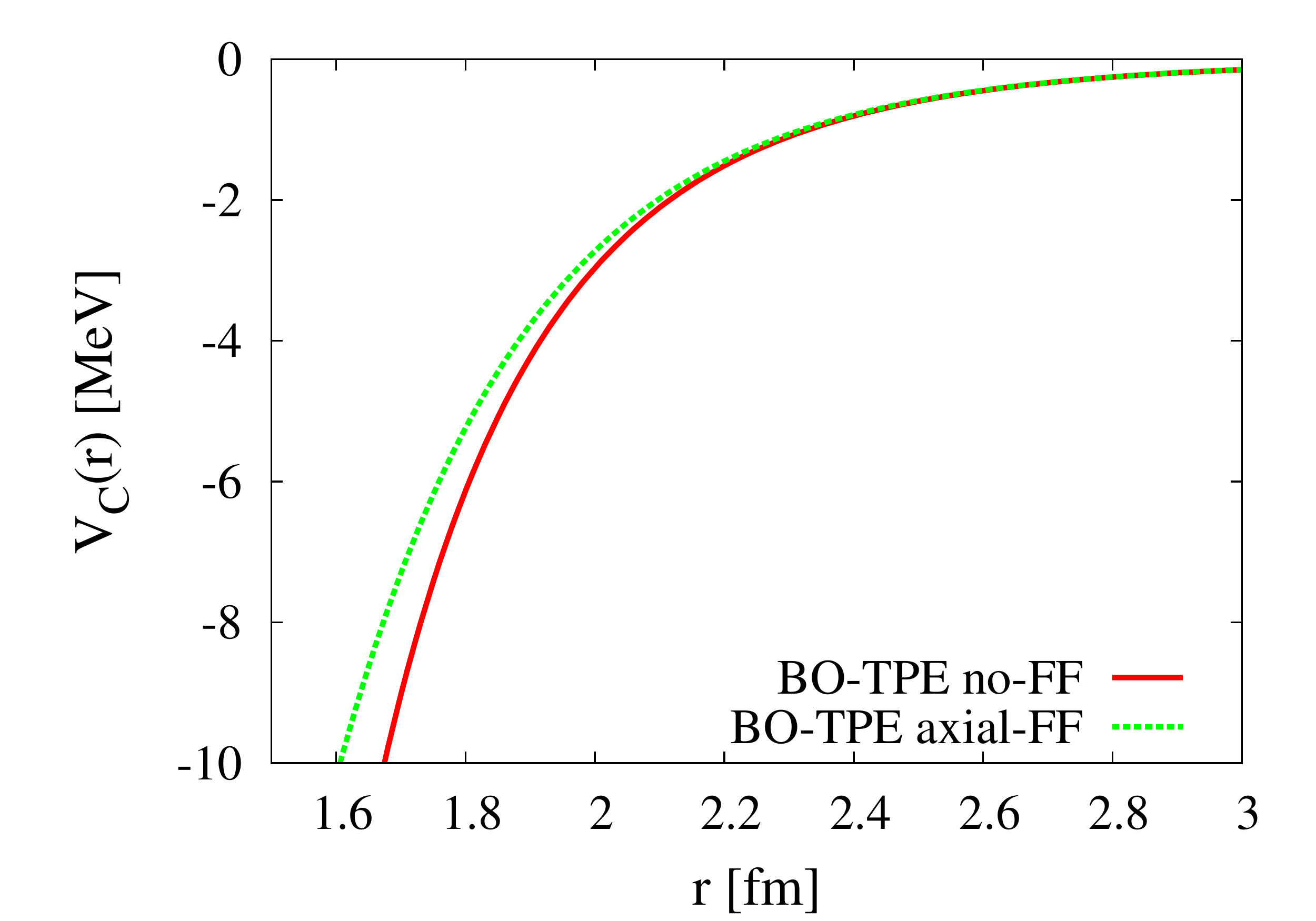} 
\includegraphics[height=4.1cm,width=4cm,angle=-90]{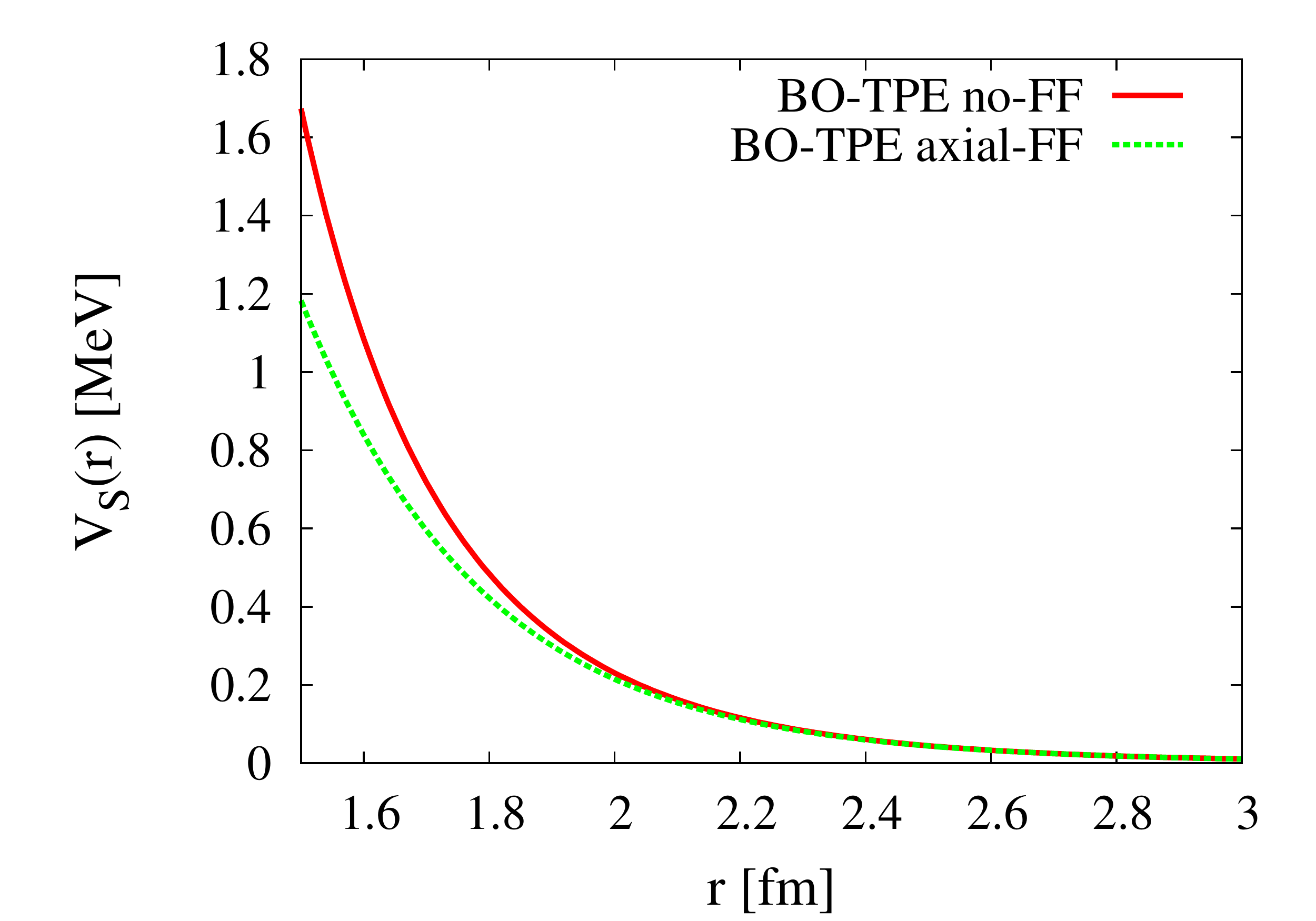} 
\includegraphics[height=4.1cm,width=4cm,angle=-90]{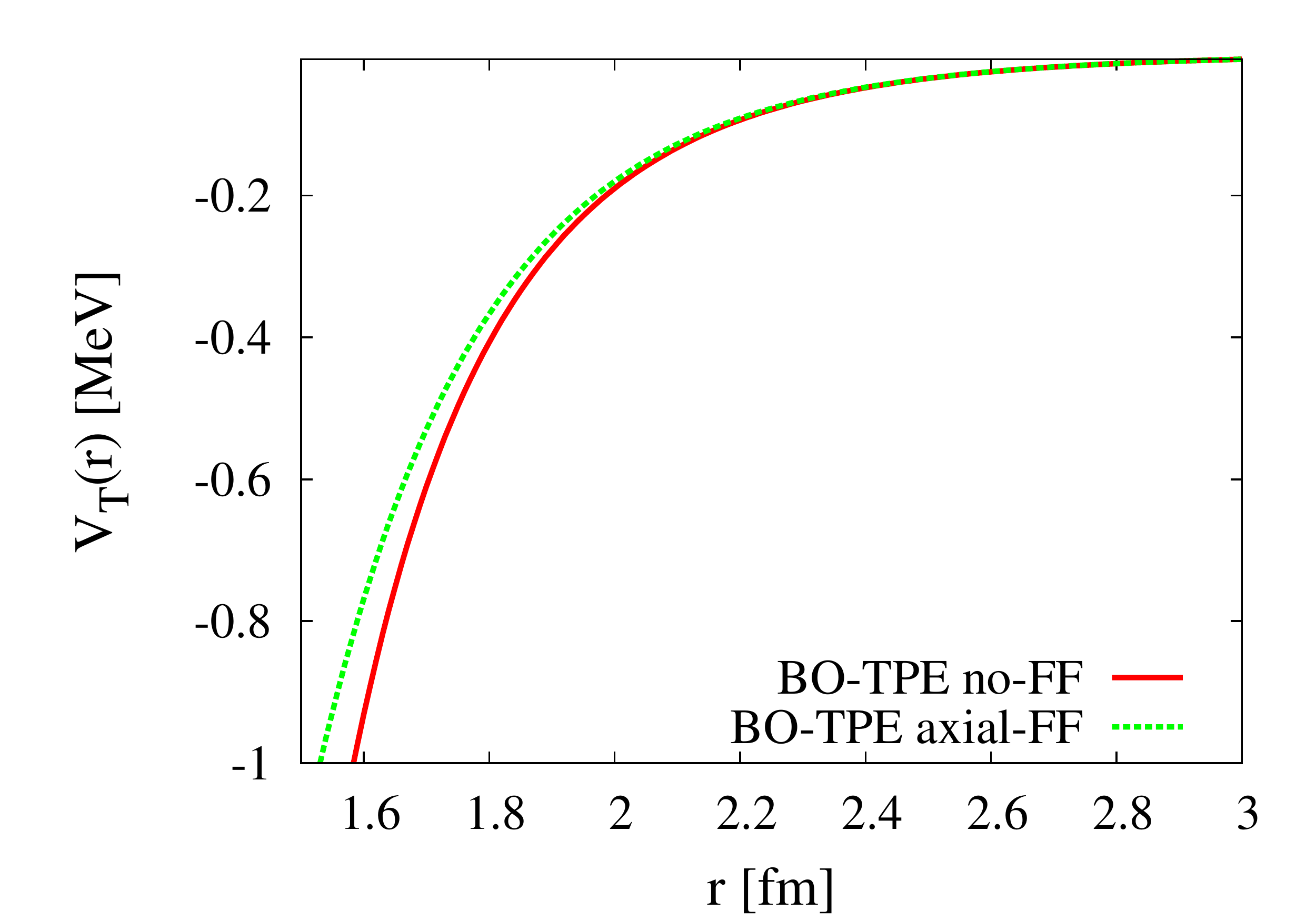} \\
\includegraphics[height=4.1cm,width=4cm,angle=-90]{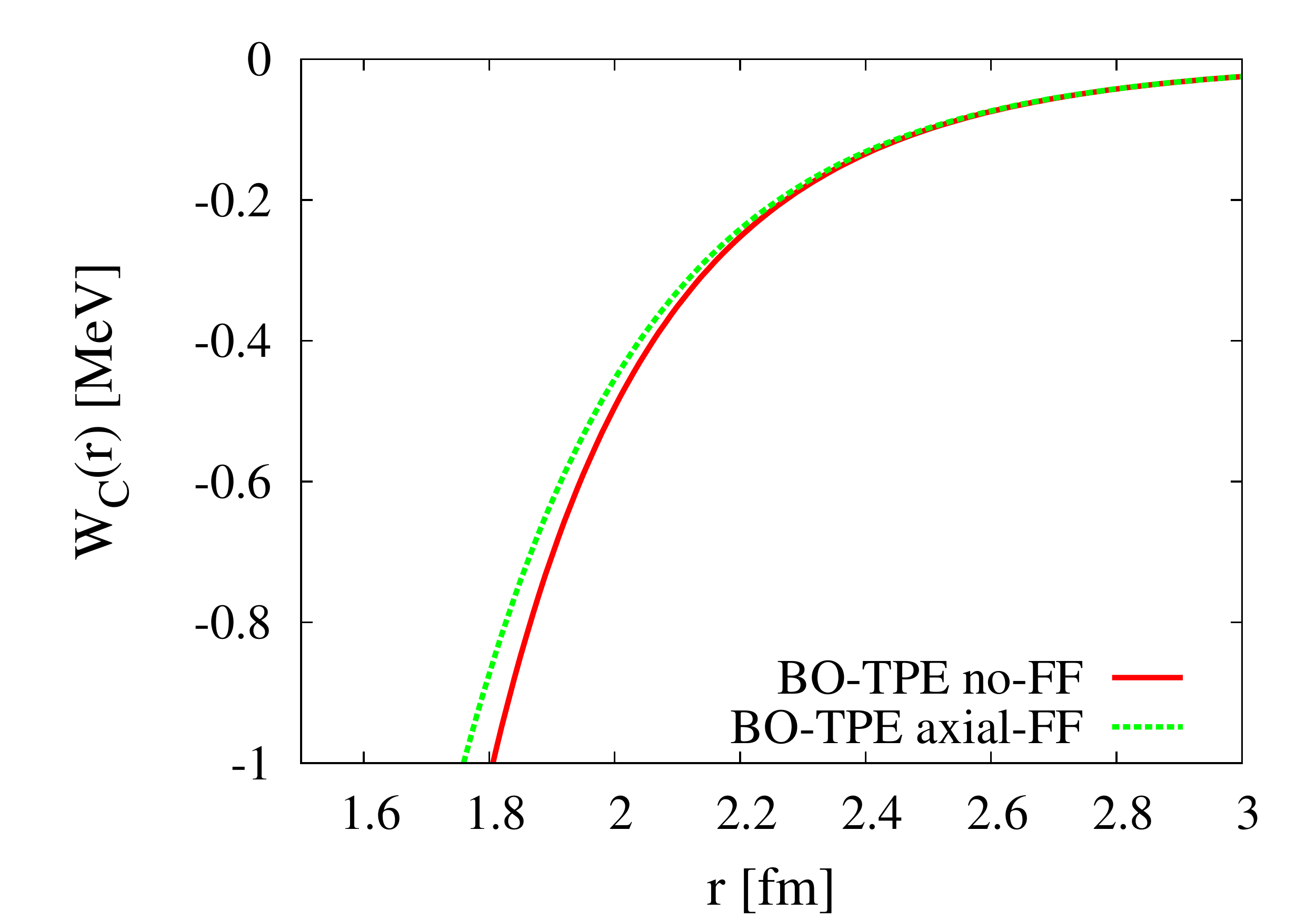} 
\includegraphics[height=4.1cm,width=4cm,angle=-90]{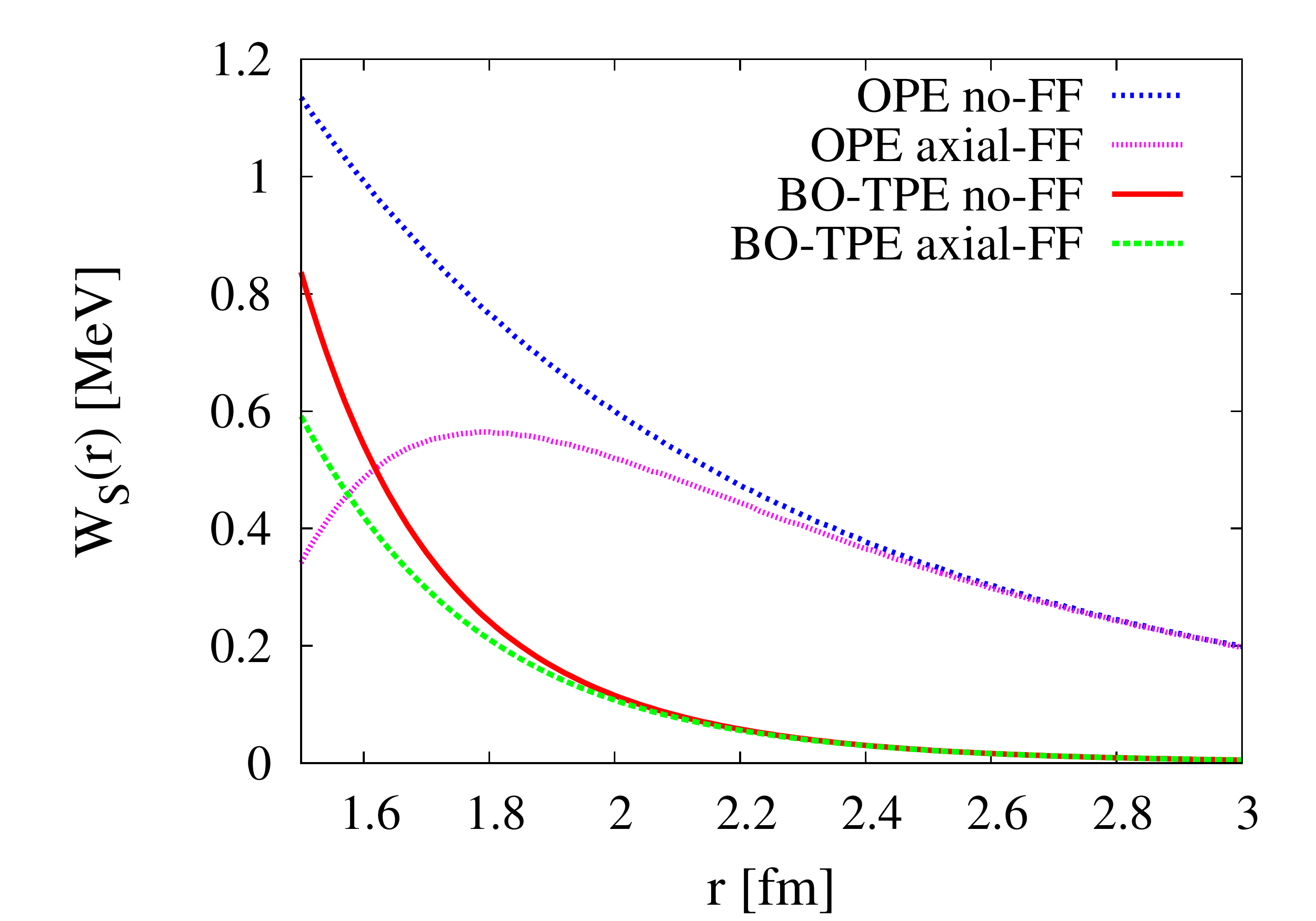} 
\includegraphics[height=4.1cm,width=4cm,angle=-90]{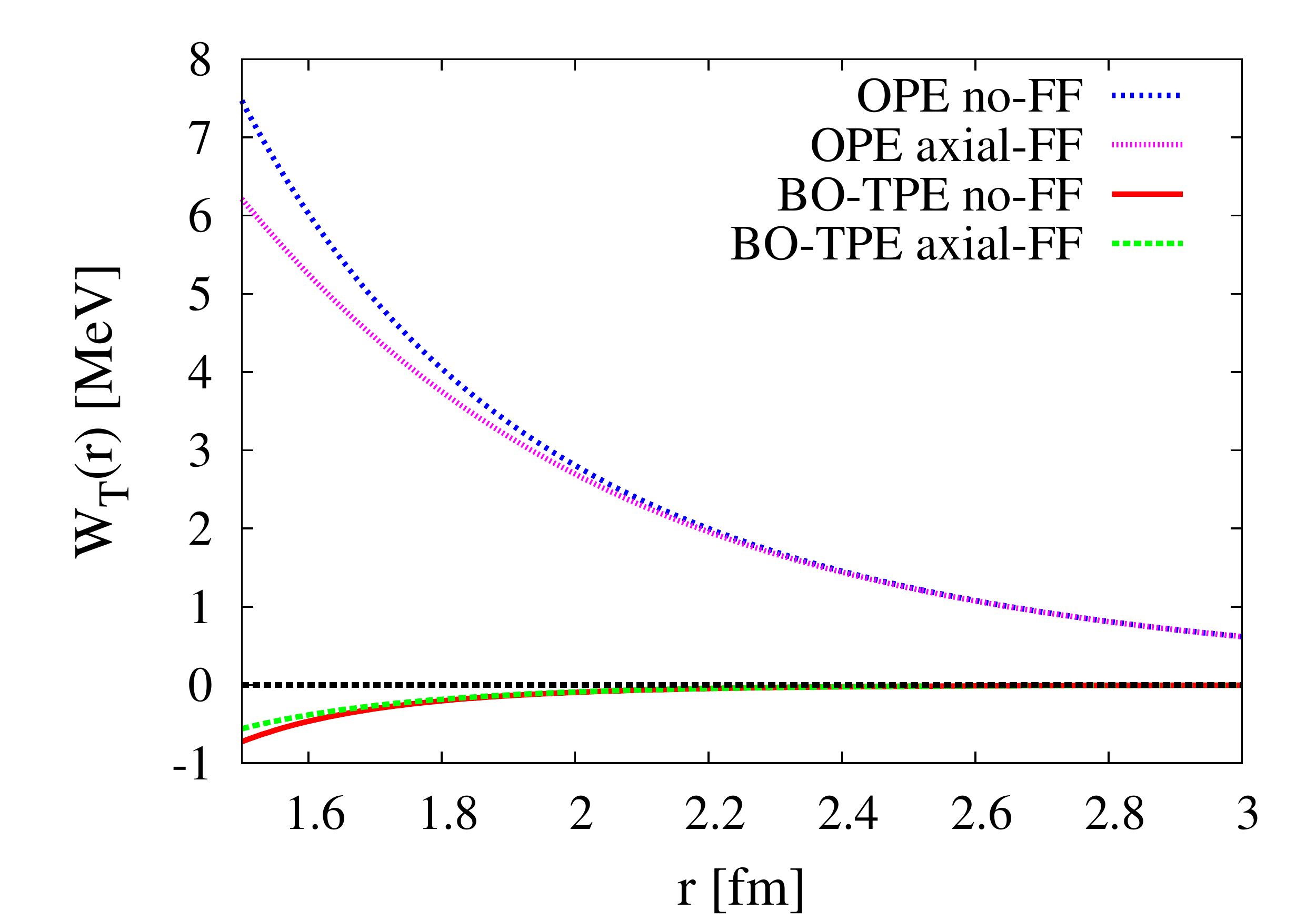} 
\end{center}
\caption{NN Potentials in the Born-Oppenheimer, including OPE and TPE
  via $\Delta$ excitation, as a function of NN distance. We show the
  effect of axial-vector (transition) form factors, generating an
  axial screening of the interaction below the elementarity radius $
  r_e \sim {\rm 2 fm}$.}
\label{Fig:born-oppen}
\end{figure}

The elementarity size $r_e$ of the nucleon can operationally be
characterized by looking at departures from point-like behaviour. If
we look for instance at the electromagnetic interaction,
Fig.~\ref{Fig:Vlong} (left panel), it clearly suggests that protons
interact as point-like charges, $1/r$, above $r> r_e =2{\rm fm}$. The
electric charge screening feature we see in the electromagnetic case
also holds for the strong interaction but here, it corresponds to
axial charge screening. This can be illustrated within a Quark Cluster
picture using a Chiral Quark
model~\cite{RuizArriola:2009vp,Cordon:2011yd}. At long distances one
can determine the NN potential in the Born-Oppenheimer approximation,
allowing for $\Delta$ isobar intermediate states, but include the $\pi
N \Delta $ transition form factor. In the Chiral Quark model, where
PCAC holds the vertices $\pi NN$, $\pi N \Delta$ and $\pi \Delta
\Delta$ are proportional to the Axial Form Factors, for which we may
use axial-vector meson dominance~\cite{Masjuan:2012sk}. This provides
a dominating contribution to the Two Pion Exchange (TPE) for $r >
r_e$. The result is shown in Fig.~\ref{Fig:born-oppen} and, as we see,
the elementarity radius turns out, again, to be about $r_e=2 {\rm fm}
\sim d_{NN}$ the average distance of nucleons in nuclear matter,
suggesting that nucleon compositeness should be play a marginal role
in nuclei.

The Quark model estimate suggests that we may pin down the interaction
as if nucleons were elementary particles down to the elementarity
radius $r_e= 2 {\rm fm}$. The practical utility of this informaton is
that we can compare different contributions stemming from meson
exchange {\it above} the elementarity radius $r_e$, showing that
(unregularized) OPE is the main contribution for distances larger than
a cut-off scale $r > r_c = 3 {\rm fm}$ below which TPE starts
contributing significantly.  This pattern occurs also with the
exchange of heavier mesons including
$\sigma,\rho,\omega,\eta,\delta,a_1$ etc. which allows us to discard
them in our analysis as well as TPE. In fact, one common feature of
all the high quality interactions is that they contain unregularized
OPE above 3 fm. Thus, we will take
\begin{eqnarray}
V_{NN} (r)|_{\rm strong} = V_{NN}^{\rm OPE}(r) \, ,  \qquad r> r_c= 3 {\rm fm} \, . 
\end{eqnarray}

Until 1993 most ``high quality fits'' carried out by many groups in
Bonn, Paris Washington and Nijmegen never improved over a $\chi^2/\nu \sim
2$.  From the point of view of statistical analysis this is
undesirable; one finds the closest theory to experiment but the
difference between theory and experiment cannot be identified as a
random fluctuation, precluding any sensible error analysis. The great
achievement of the Nijmegen group in the 1990's was to pursue all
effects which could explain the very accurate pp and np
measurements. Besides Coulomb this includes three main effects: vacuum
polarization (VP), relativistic corrections and magnetic moments (MM)
interactions. For instance VP dominates over MM for $r < 1000$fm
whereas beyond the effect is the opposite, see
Fig.~\ref{Fig:Vlong}. The em piece becomes,
\begin{eqnarray}
V_{NN} (r)|_{\rm em} = V_{NN}^{C}(r) + V_{NN}^{MM}(r) + V_{NN}^{VP}(r)+ V_{Rel}^{C}(r) \, ,  \qquad r> r_c= 3 {\rm fm} \, . 
\end{eqnarray}

As mentioned before we cannot pin down details below $\Delta r =
1/p_{\rm max}$, so we explicitly coarse grain the interaction in the
innermost unknown region, which means in practice sampling the
``unknown'' original potential at equidistant values separated by a
distance $\Delta r=0.6 {\rm fm}$, $r_n=n \Delta r$, using Dirac
delta-shells. This was an idea suggested by Aviles long
ago~\cite{Aviles:1973ee} and rediscovered recently~\cite{Entem:2007jg}
which means
\begin{eqnarray}
V(r)|_{\rm strong} = V(r)_{\rm DS} \equiv \Delta r  \sum_{n=1}^5 V_{\Delta r}(r_n) \delta (r-r_n) \qquad r \le
r_c = {\rm fm}
\end{eqnarray}
where $V_{\Delta r}(r_n)$ become the fitting parameters which depend
on the resolution scale $\Delta r$ and whose total number can be
estimated {\it a priori}~\cite{Perez:2013cza}. This can be seen by
inspecting Fig.~\ref{Fig:TLAB-complex} (right panel) by counting how
many partial waves and how many points $r_n$ per partial wave sample
the interaction with resolution $\Delta r$. This viewpoint provides a
rationale for the number of parameters, $N_{\rm Par}$, (typically
about 40-50) that were traditionally needed in the past for high
quality fits and allows at the same time the most succesful pp and np
fit to date.

\section{Fitting and selecting data}

Fitting and selecting data are intertwined, particularly when there
are incompatible data, as it is the case in NN scattering. We have
collected $N_{\rm Dat}=8000$ np+pp scattering data, $O_i^{\rm exp}$,
measured between 1950 and 2013 below $E_{\rm LAB}=350 {\rm MeV}$, with
given experimental uncertainty $\Delta O_i$. On the other hand the NN
potential
\begin{eqnarray}
V_{\rm NN}(r) = 
V_{\rm DS} (r) \theta (r_c-r) + 
\left\{ V_{\rm NN}(r)|_{\rm OPE} + V_{\rm NN}(r)|_{\rm em} \right\} \theta(r-r_c)
\end{eqnarray}
will generate $O_i^{\rm th}$, and we can
minimize the distance between theory and experiment by tuning the
fitting parameters, the potential at the coarse grained distances,
$[V(r_n)]_{l,l'}^{JS}$, and the pion-nucleon coupling constants
$f_0^2,f_p^2,f_c^2$.

Even in the case of mutually consistent data, we can never be sure
that the phenomenological theory is correct, so one poses the
classical (and admittedly twisted) statistical question as follows:
Assuming that the theory is correct, what is the probability $q$ that
the data are {\it not} described by the theory ?. This corresponds to
find the probability that for the measured observables, we {\it
  cannot} say that the true values fulfill the relation
\begin{eqnarray}
O_i=O_i^{\rm th} + \xi_i \Delta O_i \, ,  
\label{eq:generator}
\end{eqnarray}
with $i=1, \dots, N_{\rm Dat}$ and $\xi_i$ are independent random
{\it normal} variables with vanishing mean value $\langle \xi_i
\rangle =0$ and unit variance $\langle \xi_i \xi_j \rangle =
\delta_{ij}$, implying that $\langle O_i \rangle =O_i^{\rm th} $. If
this probability is large then we can discard the phenomenological
theory and look for a better one. But if it turns out to be small,
there is no good reason to discard it, and moreover we can vary the
parameters in such a way that they cover the fluctuations in the data.
The p-value is $p=1-q$ and determines the confidence level we have on
the theory. This is the standard set up for the $\chi^2$ least square
fitting, since the sum of $\nu$-gaussians, $\sum_{i=1}^\nu \xi_i^2$
has a $\chi^2$-distribution with $\nu$-degrees of freedom.  Of course,
one can only check this assumption {\it after} the optimal fit has
been carried out, and determining whether the outcoming residuals,
\begin{eqnarray}
R_i =\frac{O_i^{\rm exp}-O_i^{\rm th}}{\Delta O_i} 
\end{eqnarray}
belong to a gaussian distribution as we initially assumed. If this is
the case the fit is self-consistent. The most popular $\chi^2$-test
provides a $p$-value of $68\%$ when
\begin{eqnarray}
1 - \sqrt{2/\nu} \le \chi_{\rm min}^2/\nu \le 1 + \sqrt{2/\nu}
\end{eqnarray}
where $\nu=N_{\rm Dat}-N_{\rm Par}$. In previous works we extensively
discuss more stringent tests as well as the conditions allowing a
legitimate global scaling of
errors~\cite{Perez:2014yla,Perez:2014kpa,Perez:2015pea}.

One important issue here is the role played by the number of fitting
parameters, which we claim to be optimally fixed by the maximum
scattering LAB energy.  Obviously, if we have too few parameters a
successfull fit will not be acomplished~\footnote{Of course with
  decreasing energy the number of essential parameters decreases; for
  instance at threshold only two scattering lengths are needed for the
  only non-vanishing S-wave contributions.}. On the other hand,
although there is no limit in principle to include more parameters, we
expect strong correlations among them which display explicitly a
undesirable parameter redundancy and no real fit
improvement~\footnote{In addition, the covariance matrix size
  increases and, furthermore, may become numerically singular,
  preventing both efficiently finding an optimum and making an
  assesment of uncertainties.}. In our case we found about 50
parameters to be realistic and with moderate
correlations~\cite{Perez:2014yla,Perez:2014kpa}.

All this is fine provided we have a collection of mutually compatible
data. When this is not the case, we may ask which experiment or datum
including its error estimate is correct ${\cal O}_i^{\rm exp} \pm
\Delta {\cal O}_i$. This may not necessarily mean genuinely wrong
experiments, but rather unrealistic error estimates. Note that the
main purpose of a fit is to determine the true values of certain
parameters with a given and admissible confidence level, so we search
for a maximization of experimental consensus by excluding data sets
inconsistent with the rest of the database {\it within} the
fitting model. We  extend the Nijmegen $3\sigma$
criterion~\cite{Stoks:1993tb} by the following selection process:
\begin{enumerate}
\item Fit to all data.  If $\chi^2/\nu < 1$ you can stop. If not
  proceed further.
\item  Remove data sets with improbably high or low $\chi^2$
(3$\sigma$ criterion)
\item  Refit parameters for the remaining data. 
\item   Re-apply $3\sigma$ criterion to all data
\item   Repeat until no more data are excluded or recovered
\end{enumerate}
The effect of the selection criterion is to go from $\chi^2/\nu|_{\rm
  all}=1.41 $ to $ \chi^2/\nu|_{\rm selected}=1.05$ with a reduction
in the number of data from $N_{\rm Data}= 8173 $ to $ N_{\rm Data}=
6713$.  While this seems a drastic rejection of data it allows to
perform the largest self-consistent fit to date below 350 MeV. For
such a large number of data this is {\it not} a minor improvement; it makes
the difference between having $p \ll 1$ or $p \sim 0.68$.

In the process of selecting and fitting we have learned some features
of the phenomenological interaction within the coarse grained aproach
where no {\it a priori} condition on the fitting parameters was
imposed.  We can fit the pp database independently. However, the
isovector phases in the pn system are largely uncontrolled by the np
data. Therefore it is preferable to fit the pp first, and to refit the
pp+np system simultaneously by making some statistically testable
isospin assumptions.
 
We have also tried to use normality of the residuals as a rejection
tool, without much success. The reason is that a normality test checks
the excess or deficit of residuals as compared to the guess
distribution, but does not indicate specifically which data are
responsible for normality deviations.

Likewise, we have analyzed the robustness of the database with respect
to restricting or enlarging the rejection level from $3\sigma$ to $2
\sigma$ or $4\sigma$, respectively. In the first case, it was not
possible to find a self-consistent database with the number of
accepted data fluctuating from subsequent fits. This is probably
related to the grouping of data in a common experiment, preventing a
stable decision of accepting/rejecting data groups. In the
$4\sigma$-rejection level case, the self-consistent database exists
but does not comply to the normality test at the imposed significance
level.

Why are these apparently small details important?. One reason is that
nuclear structure calculations are insensitive to long range
potential, but quite dependent on the least known mid range part, so
that errors are propagated to binding energies or matrix
elements~\cite{NavarroPerez:2012vr,Perez:2012kt,Amaro:2013zka,Perez:2014laa,Perez:2015bqa,Perez:2016oia}. The
role played by $\chi$TPE and determination of chiral constants $c_1$,
$c_3$ $c_4$~\cite{Perez:2013oba,Perez:2014bua} (where $N_{\rm Par}=30$
and $r_c = 1.8 {\rm}$) and inclusion $\Delta
$-resonance~\cite{Piarulli:2014bda} have been analyzed with the
Granada-2013 database. .

In all, the present situation regarding both the selection of data and
the normality of residuals is highly satisfactory. In our view, this
combined consistency of the statistical assumptions and the model
analyzing them provides a good starting point to proceed further to
determine the pion-nucleon coupling constants.

\section{Determination of $\pi$NN coupling constants}

The charge symmetry breaking is restricted to mass differences by
setting $f_{p} = -f_{n} = f_c = f$ and the value $f^2|_{\rm Nij} =
0.0750(9)$ recommended by the Nijmegen group~\cite{Bergervoet:1987tr}
has been used in most of the potentials since the seminal 1993 partial
wave analysis~\cite{Stoks:1993tb}.  In their 1997 status
report\cite{deSwart:1997ep} the Nijmegen group wrote: {\it ``The
  present accuracies in the determination of the various coupling
  constants are such, that with a little improvement in the data and
  in the analyses these charge-independence breaking effects could be
  checked.''}. The Granada-2013 database has 6713 data compared to the
4313 of Nijmegen-1993. Can this be the invoked little improvement ?.

We try to answer this by recalling that electroweak
corrections scale with the fine structure constant $\alpha=1/137$ and
the light quark mass differences. Thus
\begin{eqnarray}
\frac{\delta g}{g} = {\cal O} \left(\alpha, \frac{m_u-m_d}{\Lambda_{\rm QCD}} \right) = {\cal O} \left(\alpha, \frac{M_p-M_n}{\Lambda_{\rm QCD}} \right) 
\end{eqnarray}
for the relative change around a mean value. These are naturally at
the $1-2\%$ level, a small effect. The question is on how many
independent measurements are needed to achieve this desired
accuracy. According to the central limit theorem, for $N$ direct
independent measurements the relative standard deviation scales as
$$
\frac{\Delta g}{g} = {\cal O} \left( \frac{1}{\sqrt{N}} 
\right) 
$$ and $\delta g \sim \Delta g$ for $N=7000-10000$. We cannot carry
out these direct measurements of $g$ but we can proceed indirectly by
considering a set of mutually consistent NN scattering measurements $
O_i^{\rm exp}$ with $i=1, \dots, N_{\rm Dat}$ and use a model with $g$
and $ \vec \lambda =(\lambda_1, \dots , \lambda_{N_{\rm par}})$
parameters, which produces $O^{\rm th}_i (g, \vec \lambda) $. We can
then eliminate the parameters $\vec \lambda$, in favor of $N_{\rm Par}$
experiments and we are left with $N=N_{\rm Dat}-N_{\rm Par}$
independent observables which depend just on $g$ providing $N$
independent determinations. Of course, these measurements will have
some statistical error, so that $O_i = \bar O_i \pm \Delta O_i$, which
means that $O_i$ is a random variable, and a $\chi^2$-fit is nothing
but a democratic way of eliminating the parameters. Since $N_{\rm Dat}
\gg N_{\rm Par}$, we need about $N_{\rm Dat}\sim 7000-10000$ to
witness isospin breaking with the coarse grained interaction, and in
our recent work we do~\cite{Perez:2016aol}. From our full covariance
matrix analysis we get for the $g's$
\begin{eqnarray}
g_n^2/(4\pi)= 14.91(39) \, , \qquad g_c^2/(4\pi)=13.81(11) , \qquad
g_p^2/(4\pi)=13.72(7) \, .  
\label{eq:g's}
\end{eqnarray}
We thus confirm the premonition of the Nijmegen group, although
further ``little improvements'' are still needed to confirm an
ordering pattern, such as e.g. Eq.~(\ref{eq:qm-order}).

\begin{figure}[hpt]
\begin{center}
\epsfig{figure=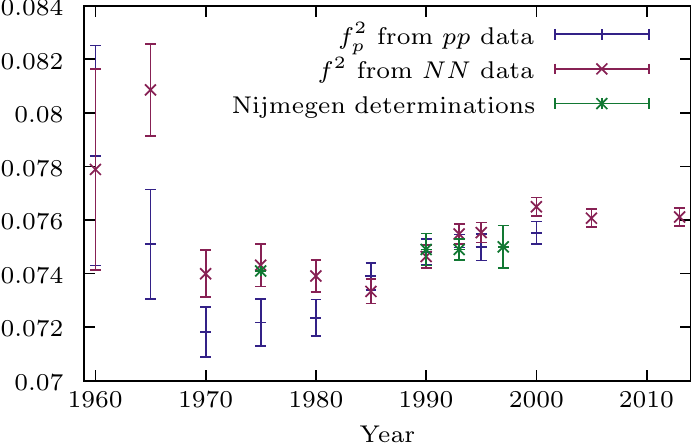,width=0.7\linewidth}  
\end{center}
\caption{(Color online) Chronological recreation of pion-nucleon
  coupling constants determinations from NN data compared to several Nijmegen determinations.(see main text)}
\label{Fig:f2byyear}
\end{figure}
In Fig.~\ref{Fig:f2byyear} we show a chronological recreation of
$f_p^2$ and $f^2$ determinations using at any rate the NN data of the
complete database measured up to a given year, which expectedly
resembles the historic plot~\cite{Sainio:1999ba}.  We consider in any
case and, when needed, the corresponding $3\sigma$-consistent
database. We also plot several Nijmegen determinations including
$f^2|_{\rm Nij}$~\cite{deSwart:1997ep}. Assuming a unique pion-nucleon
coupling constant we obtain $f^2=0.07611(33)$ which is $1\sigma$
compatible but almost three times more accurate. The latest most
accurate $\pi N$ scattering determinations
has are based on the GMO rule~\cite{Ericson:2000md} with 
$g_c^2/(4\pi)=14.11(20)$, use fixed-t dispersion relations~\cite{Arndt:2006bf}
with $g_c^2/(4\pi)=13.76(8)$ and are based on $\pi N$
scattering lengths and $\pi^- d$ scattering and the GMO sum rule
yielding $g_c^2/(4\pi)=13.69(19)$~\cite{Baru:2010xn}. Our value,
Eq.~(\ref{eq:g's}) is compatible with this last determination, but
twice more accurate.

\section{Conclusions}

According to our analysis neutrons interact more strongly than protons
above $r_c=3 {\rm fm}$, but we cannot check what is the influence on
neutron-neutron scattering as we have not determined the
nn-interaction {\it below} $r_c$. The traditional and compelling
explanation that the nn scatering length is {\it larger} than the
strong contribution to the pp-scattering length (a model dependent
quantity) would rest on the assumption that there is no relevant
isospin breaking below 3 fm, a fact that is not supported by our
analysis and requires further understanding. 

Isospin breaking at short distances has always been a difficult
subject.  We suggest to cut the gordian knot by separating the NN
interaction in two distinct regions marked by a short distance radius
$r_c$ and assuming a charge dependent one pion exchange above
$r_c$, as the unique strong contribution. Below this radius we
purposely ignore the specific form of the interaction by coarse
graining it down to $\Delta r=0.6 {\rm fm}$, the shortest deBroglie
wavelength before pions are produced. This distance turns out to be
$r_c=3 {\rm fm}$ and we can describe 6727 NN sattering data with a
total $\chi^2=6907$ and 55 short distance parameters plus $f_p^2,f_0^2$
and $f_c^2$.
%three
%pion-nucleon coupling constants $f_p^2 = 0.0759(4)$, $f_{0}^2
%=0.079(1)$ and $f_{c}^2 = 0.0763(6)$.

\section*{Acknowledgments}

One of us (E.R.A.) would like to thank J.L. Goity and J. Ruiz de
Elvira for discussions, the Mainz Institute for Theoretical Physics
(MITP) for its hospitality and support and the organizers I. Caprini,
K. Chetyrkin, C. Dominguez, A. Pich and H. Spiesberger for the
invitation and the nice workshop atmosphere.

%\section*{References}

%\bibliographystyle{ws-ijmpe} 
%\bibliography{../Paper/fpiNN-deltashell}

\begin{thebibliography}{10}

\bibitem{Yukawa:1935xg}
H.~Yukawa, {\em Proc. Phys. Math. Soc. Jap.} {\bf 17}  (1935) 48.
%, [Prog. Theor.  Phys. Suppl.1,1(1935)].

\bibitem{Kemmer:1939zz}
N.~Kemmer, {\em Proc. Roy. Soc. Lond.} {\bf A173}  (1939) 91.

\bibitem{pauli1948meson}
W.~Pauli, {\em Meson theory of nuclear forces} (Interscience Publishers, 1948).


\bibitem{Bethe:1940zz}
H.~Bethe, {\em Phys. Rev.} {\bf 57} (1940) 260, {\em Phys.Rev.} {\bf 57}  (1940) 390.

\bibitem{deSwart:1997ep}
J.~de~Swart, M.~Rentmeester and R.~Timmermans, {\em PiN Newslett.} {\bf 13}
  (1997) 96
%, \href{http://arxiv.org/abs/nucl-th/9802084}{{\ttfamily  arXiv:nucl-th/9802084 [nucl-th]}}.

\bibitem{Sainio:1999ba}
M.~Sainio, {\em PiN Newslett.} {\bf 15}  (1999) 156.
%  \href{http://arxiv.org/abs/hep-ph/9912337}{{\ttfamily arXiv:hep-ph/9912337  [hep-ph]}}.

\bibitem{henley1969isospin} 
E.~Henley and D.~Wilkinson, {\em Isospin
  in nuclear physics} (North-Holland 1969).

\bibitem{Miller:1990iz}
G.~A. Miller, B.~M.~K. Nefkens and I.~Slaus, {\em Phys. Rept.} {\bf 194}
  (1990) 1.

\bibitem{Miller:2006tv}
G.~A. Miller, A.~K. Opper and E.~J. Stephenson, {\em Ann.Rev.Nucl.Part.
  Sci.}{\bf 56}(2006)253.
%  \href{http://arxiv.org/abs/nucl-ex/0602021}{{\ttfamily arXiv:nucl-ex/0602021  [nucl-ex]}}.

\bibitem{Dumbrajs:1983jd}
O.~Dumbrajs {\em et~al.}, {\em
  Nucl.Phys.} {\bf B216}  (1983) 277.

\bibitem{signell1969nuclear}
P.~Signell, {\em The nuclear potential} (Springer, 1969).

\bibitem{Aoki:2011ep} S.~Aoki, {\em
  Prog.Part.Nucl.Phys.} {\bf 66}  (2011) 687.
%  \href{http://arxiv.org/abs/1107.1284}{{\ttfamily arXiv:1107.1284 [hep-lat]}}.

\bibitem{Aoki:2013tba}
S.~Aoki, {\em Eur.Phys.J.} {\bf A49}  (2013)  ~81.
%  \href{http://arxiv.org/abs/1309.4150}{{\ttfamily arXiv:1309.4150 [hep-lat]}}.

\bibitem{Alexandrou:2007xj}
C.~Alexandrou {\it et al.}
%, G.~Koutsou, T.~Leontiou, J.~W. Negele and A.~Tsapalis, {\em
  Phys.Rev. {\bf D76}  (2007)   094511.
%  \href{http://arxiv.org/abs/0912.0394}{{\ttfamily arXiv:0912.0394 [hep-lat]}}.

\bibitem{Erkol:2008yj}
G.~Erkol, M.~Oka and T.~T. Takahashi, {\em Phys.Rev.} {\bf D79}  (2009)
  074509.
%, \href{http://arxiv.org/abs/0805.3068}{{\ttfamily arXiv:0805.3068  [hep-lat]}}.

\bibitem{Aoki:2009ji}
S.~Aoki, T.~Hatsuda and N.~Ishii, {\em Prog.Theor.Phys.} {\bf 123}  (2010) 89.
%, \href{http://arxiv.org/abs/0909.5585}{{\ttfamily arXiv:0909.5585 [hep-lat]}}.

\bibitem{Manohar:1983md}
A.~Manohar and H.~Georgi, {\em Nucl. Phys.} {\bf B234}  (1984) 189.

\bibitem{RuizArriola:2002wr}
E.~Ruiz~Arriola, {\em Acta Phys. Polon.} {\bf B33}  (2002) 4443.
%  \href{http://arxiv.org/abs/hep-ph/0210007}{{\ttfamily arXiv:hep-ph/0210007
%  [hep-ph]}}.

%\cite{PavonValderrama:2005ku}
\bibitem{PavonValderrama:2005ku} 
  M.~Pavon Valderrama and E.~Ruiz Arriola,
  %``Low-energy NN scattering at next-to-next-to-next-to-next-to-leading order for partial waves with j [] 5,''
  Phys.\ Rev.\ C {\bf 72}, 044007 (2005).
%  doi:10.1103/PhysRevC.72.044007
  %%CITATION = doi:10.1103/PhysRevC.72.044007;%%
  %31 citations counted in INSPIRE as of 01 Jun 2016


\bibitem{Oller:2014uxa}
J.~A. Oller, {\em Phys. Rev.} {\bf C93}  (2016)   024002.
%  \href{http://arxiv.org/abs/1402.2449}{{\ttfamily arXiv:1402.2449 [nucl-th]}}.

\bibitem{PavonValderrama:2005gu}
M.~Pavon~Valderrama and E.~Ruiz~Arriola, {\em Phys. Rev.} {\bf C72}  (2005)
  054002.
% \href{http://arxiv.org/abs/nucl-th/0504067}{{\ttfamily  arXiv:nucl-th/0504067 [nucl-th]}}.

\bibitem{NavarroPerez:2012vr}
R.~Navarro~Perez, J.~E. Amaro and E.~Ruiz~Arriola  (2012) arXiv:1202.6624 [nucl-th].

\bibitem{Perez:2012kt}
R.~Navarro~Perez, J.~E. Amaro and E.~Ruiz~Arriola, {\em PoS} {\bf QNP2012}
  (2012)   145.
%, \href{http://arxiv.org/abs/1206.3508}{{\ttfamily  arXiv:1206.3508 [nucl-th]}}.

\bibitem{Amaro:2013zka}
J.~E. Amaro, R.~Navarro~Perez and E.~Ruiz~Arriola, {\em Few Body Syst.} {\bf
  55}  (2014) 977.
%, \href{http://arxiv.org/abs/1310.7456}{{\ttfamily  arXiv:1310.7456 [nucl-th]}}.


\bibitem{RuizArriola:2009vp} E.~Ruiz~Arriola and A.~Calle~Cordon. In
  {\em Problems in multi-quark states}. Proceedings, Mini-Workshop,
      Bled, Slovenia, June 29-July 6, (2009). arXiv:0910.1333 [hep-ph]. 
%, { {Renormalization and Universality of NN
%  interactions in Chiral Quark and Soliton Models}}, in {\em {Problems in
%  multi-quark states. Proceedings, Mini-Workshop, Bled, Slovenia, June 29-July
%  6, 2009}\/},  (2009).
%\newblock \href{http://arxiv.org/abs/0910.1333}{{\ttfamily arXiv:0910.1333
%  [hep-ph]}}.

\bibitem{Cordon:2011yd}
A.~Calle~Cordon and E.~Ruiz~Arriola. arXiv:1108.5992 [nucl-th]
%  \href{http://arxiv.org/abs/1108.5992}{{\ttfamily arXiv:1108.5992 [nucl-th]}}.

\bibitem{Masjuan:2012sk}
P.~Masjuan, E.~Ruiz~Arriola and W.~Broniowski, {\em Phys. Rev.} {\bf D87}
  (2013)   014005.
%, \href{http://arxiv.org/abs/1210.0760}{{\ttfamily  arXiv:1210.0760 [hep-ph]}}.

\bibitem{Aviles:1973ee}
J.~B. Aviles, {\em Phys. Rev.} {\bf C6}  (1972) 1467.

%\cite{Entem:2007jg}
\bibitem{Entem:2007jg}
  D.~R.~Entem, E.~Ruiz Arriola, M.~Pavon Valderrama and R.~Machleidt,
  %``Renormalization of chiral two-pion exchange NN interactions. momentum versus coordinate space,''
  Phys.\ Rev.\ C {\bf 77} (2008) 044006
%  doi:10.1103/PhysRevC.77.044006
%  [arXiv:0709.2770 [nucl-th]].
  %%CITATION = doi:10.1103/PhysRevC.77.044006;%%
  %70 citations counted in INSPIRE as of 05 Jun 2016


\bibitem{Perez:2013cza}
R.~Navarro~Perez, J.~E. Amaro and E.~Ruiz~Arriola, {\em Few Body Syst.} {\bf
  55}  (2014) 983.
%, \href{http://arxiv.org/abs/1310.8167}{{\ttfamily  arXiv:1310.8167 [nucl-th]}}.

\bibitem{Perez:2014yla}
R.~Navarro~Perez, J.~E. Amaro and E.~Ruiz~Arriola, {\em Phys. Rev.} {\bf C89}
  (2014)   064006.
%, \href{http://arxiv.org/abs/1404.0314}{{\ttfamily  arXiv:1404.0314 [nucl-th]}}.

\bibitem{Perez:2014kpa}
R.~Navarro~Perez, J.~E. Amaro and E.~Ruiz~Arriola, {\em J. Phys.} {\bf G42}
  (2015)   034013.
%, \href{http://arxiv.org/abs/1406.0625}{{\ttfamily  arXiv:1406.0625 [nucl-th]}}.

\bibitem{Perez:2015pea}
R.~Navarro~Perez, E.~Ruiz~Arriola and J.~Ruiz~de Elvira, {\em Phys. Rev.} {\bf
  D91}  (2015)   074014.
%, \href{http://arxiv.org/abs/1502.03361}{{\ttfamily  arXiv:1502.03361 [hep-ph]}}.



%\cite{Perez:2014laa}
\bibitem{Perez:2014laa} 
  R.~Navarro~Perez, E.~Garrido, J.~E.~Amaro and E.~Ruiz~Arriola,
  %``Triton binding energy with realistic statistical uncertainties,''
  Phys.\ Rev.\ C {\bf 90}, no. 4, 047001 (2014)
%  doi:10.1103/PhysRevC.90.047001
%  [arXiv:1407.7784 [nucl-th]].
  %%CITATION = doi:10.1103/PhysRevC.90.047001;%%
  %7 citations counted in INSPIRE as of 05 Jun 2016


%\cite{Perez:2015bqa}
\bibitem{Perez:2015bqa} 
  R.~Navarro Perez, J.~E.~Amaro, E.~Ruiz Arriola, P.~Maris and J.~P.~Vary,
  %``Statistical error propagation in ab initio no-core full configuration calculations of light nuclei,''
  Phys.\ Rev.\ C {\bf 92}, no. 6, 064003 (2015)
%  doi:10.1103/PhysRevC.92.064003
%  [arXiv:1510.02544 [nucl-th]].
  %%CITATION = doi:10.1103/PhysRevC.92.064003;%%
  %3 citations counted in INSPIRE as of 05 Jun 2016

%\cite{Perez:2016oia}
\bibitem{Perez:2016oia}
  R.~N.~Perez, A.~Nogga, J.~E.~Amaro and E.~R.~Arriola,
  %``Binding in light nuclei: Statistical NN uncertainties vs Computational accuracy,''
  arXiv:1604.00968 [nucl-th].
  %%CITATION = ARXIV:1604.00968;%%

%\cite{Perez:2013oba}
\bibitem{Perez:2013oba}
  R.~Navarro Pérez, J.~E.~Amaro and E.~R.~Arriola,
  %``Coarse grained NN potential with Chiral Two Pion Exchange,''
  Phys.\ Rev.\ C {\bf 89} (2014) no.2,  024004
%  doi:10.1103/PhysRevC.89.024004
%  [arXiv:1310.6972 [nucl-th]].
  %%CITATION = doi:10.1103/PhysRevC.89.024004;%%
  %22 citations counted in INSPIRE as of 05 Jun 2016


%\cite{Perez:2014bua}
\bibitem{Perez:2014bua}
  R.~Navarro Perez, J.~E.~Amaro and E.~Ruiz~Arriola,
  %``Low energy chiral two pion exchange potential with statistical uncertainties,''
  Phys.\ Rev.\ C {\bf 91} (2015) no.5,  054002
%  doi:10.1103/PhysRevC.91.054002
%  [arXiv:1411.1212 [nucl-th]].
  %%CITATION = doi:10.1103/PhysRevC.91.054002;%%
  %10 citations counted in INSPIRE as of 05 Jun 2016

%\cite{Piarulli:2014bda}
\bibitem{Piarulli:2014bda}
  M.~Piarulli, L.~Girlanda, R.~Schiavilla, R.~Navarro Perez, J.~E.~Amaro and E.~Ruiz Arriola,
  %``Minimally nonlocal nucleon-nucleon potentials with chiral two-pion exchange including $\Delta$ resonances,''
  Phys.\ Rev.\ C {\bf 91} (2015) no.2,  024003
%  doi:10.1103/PhysRevC.91.024003
%  [arXiv:1412.6446 [nucl-th]].
  %%CITATION = doi:10.1103/PhysRevC.91.024003;%%
  %15 citations counted in INSPIRE as of 05 Jun 2016



\bibitem{Bergervoet:1987tr}
J.~Bergervoet, P.~van Campen, T.~Rijken and J.~de~Swart, {\em Phys.Rev.Lett.}
  {\bf 59}  (1987)   2255.

\bibitem{Stoks:1993tb}
V.~Stoks, R.~Kompl, M.~Rentmeester and J.~de~Swart, {\em Phys.Rev.} {\bf C48}
  (1993) 792.

\bibitem{Perez:2016aol}
R.~Navarro~Perez, J.~E. Amaro and E.~Ruiz~Arriola  (2016). arXiv:1606.00592
  [nucl-th]
%  \href{http://arxiv.org/abs/1606.00592}{{\ttfamily }}.

%\cite{Ericson:2000md}
\bibitem{Ericson:2000md} 
  T.~E.~O.~Ericson, B.~Loiseau and A.~W.~Thomas,
  %``Determination of the pion nucleon coupling constant and scattering lengths,''
  Phys.\ Rev.\ C {\bf 66}, 014005 (2002)
%  doi:10.1103/PhysRevC.66.014005
%  [hep-ph/0009312].
  %%CITATION = doi:10.1103/PhysRevC.66.014005;%%
  %114 citations counted in INSPIRE as of 05 Jun 2016

%\cite{Arndt:2006bf}
\bibitem{Arndt:2006bf} 
  R.~A.~Arndt, W.~J.~Briscoe, I.~I.~Strakovsky and R.~L.~Workman,
  %``Extended partial-wave analysis of piN scattering data,''
  Phys.\ Rev.\ C {\bf 74}, 045205 (2006)
%  doi:10.1103/PhysRevC.74.045205
%  [nucl-th/0605082].
  %%CITATION = doi:10.1103/PhysRevC.74.045205;%%
  %299 citations counted in INSPIRE as of 05 Jun 2016


\bibitem{Baru:2010xn}
V.~Baru, C.~Hanhart, M.~Hoferichter, B.~Kubis, A.~Nogga {\em et~al.}, {\em
  Phys.Lett.} {\bf B694}  (2011) 473. 
%  \href{http://arxiv.org/abs/1003.4444}{{\ttfamily arXiv:1003.4444 [nucl-th]}}.




\end{thebibliography}

\end{document}